\documentclass{article}
\usepackage{graphics}

\catcode`\@=11
\ifcase \@ptsize
\or 
\or 
\fi
\advance\textheight by \topskip
\catcode`\@=12

\begin{document}

\title{\Large \bf The Superfluid State of a Bose Liquid
as a Superposition of a Single-Particle and Pair Coherent
Condensates}

\author {\large E.A.~Pashitskij$^1$, S.V.~Mashkevich$^2$,
S.I.Vilchynskyy$^3$\\
$^1$\it Institute for Physics, NAS of Ukraine, Kiev 03022, Ukraine \\
\it pashitsk@iop.kiev.ua\\
$^2$\it Institute for Theoretical Physics, Kiev 03143,
Ukraine\footnote{On leave of absence.} \\
\it mash@mashke.org\\
$^3$\it T. Shevchenko Kiev University, Kiev 03022, Ukraine \\
\it sivil@ap3.bitp.kiev.ua }
\maketitle
\begin{abstract}

One considers the superfluid (SF) state of a Bose liquid with a
strong repulsion between bosons, in which at $T=0$, along with a
weak single-particle Bose-Einstein condensate (BEC), there exists
an intensive pair coherent condensate (PCC), analogous to the
Cooper condensate in a Fermi liquid with an attraction between
the fermions. Such a PCC emerges in a system of bosons due to an
oscillating sign-changing momentum dependence of the Fourier
component of the pair interaction potential, which is characteristic
of a certain family of repulsion potentials, for example, for a
regularized ``hard spheres'' model or for potentials with finite
jumps or inflection points with an infinite derivative.
In such cases, the Fourier component is negative in some domain of
nonzero momentum transfer, which corresponds to an effective
attraction of a quantum mechanical (diffraction) nature.
The collective effects of renormalization (``screening'') of the
initial interaction, which are described by the bosonic polarization
operator, due to its negative sign on the ``mass shell'', lead to
a suppression of the repulsion and an enhancement of the effective
attraction in the respective domains of momentum space.
In the process of building a self-consistent model of the SF state,
it is the ratio of the BEC density to the full density of the liquid
$n_0/n\ll 1$ that is used as a small parameter---unlike in the
Bogolyubov theory for a quasi-ideal Bose gas, in which the small
parameter is the ratio of the number of supracondensate excitations
to the number of particles in an intensive BEC, $(n-n_0)/n_0\ll 1$.
A closed system of nonlinear integral equations for the normal
$\tilde\Sigma_{11}(\mathbf{p}, \omega)$ and anomalous
$\tilde\Sigma_{12}(\mathbf{p}, \omega)$ self-energy parts is obtained,
in the framework of a renormalized perturbation theory built on
combined hydrodynamic (at  $p\to 0$) and field (at $p\ne 0 $)
variables, whose usage eliminates infrared divergencies and ensures
that the functions $\tilde\Sigma_{ij}(\mathbf{p}, \epsilon)$ are
analytic at $p\to 0 $ and $\epsilon\to 0$ and that the SF order
parameter $\tilde\Sigma_{12}(0,0)\ne 0$ at $T=0$.

In the framework of the hard-spheres model, a spectrum of quasiparticles
is obtained, which is in good accordance with the experimental spectrum
of elementary excitations in superfluid $^4$He.
It is shown that the roton minimum in the quasiparticle spectrum is
directly associated with the first negative minimum of the Fourier
component of the renormalized potential of pair interaction between
the bosons.
Finally, the question of applicability of the Landau criterion to
the description of the SF state of $^4$He in the absence of quantum
vortices is discussed.

PACS: 67.57.-z
\end{abstract}

\large
\section{Introduction}
Despite the big progress achieved in the theory of superfluidity
since the pioneer works by Landau \cite{LLD},
Bogolyubov \cite{BNN}, Feynman \cite{FR} and others
\cite{7}--\cite{36}, the task of constructing a microscopic theory of
a superfluid (SF) state of a  $^4$He Bose liquid cannot be
considered complete. In fact, such questions of principle as:
(i) the origin of the roton minimum in the spectrum of elementary
excitations; (ii) applicability of the Landau criterion of superfluidity
for the determination of the critical velocity of dissipationless
flow in SF helium (He~II); (iii) the quantum mechanical structure
of the SF component of the $^4$He Bose liquid below the $\lambda$
point, at $T<T_\lambda=2.17$~K, etc., remain unsolved.
In particular, according to the latest results in quantum evaporation
of $^4$He atoms \cite{9}, the maximal density $\rho_0$ of the
single-particle Bose-Einstein condensate (BEC) in the $^4$He
Bose liquid even at very low temperatures $T\ll T_\lambda$
does not exceed $10 \%$ of the total density $\rho$ of liquid
$^4$He, whereas the density of the SF component
$\rho_s \to \rho $ at $T\to 0$.
Such a low density of the BEC is implied by a strong interaction
between $^4$He atoms and is an indication of the fact that such an
``exhausted'' BEC cannot by itself form the microscopic basis of
the SF component $\rho_s$. Therefore, the quantum structure
of the SF condensate in He~II with the ``excess'' density
$(\rho_s-\rho_0)\gg \rho_0$ calls for a more thorough
investigation \cite{10a}--\cite{11}.

On the other hand, numerous precise experiments on the restoration
of the dynamic structure factor $S(\mathbf{p}, \epsilon)$ in
liquid $^4$He, involving inelastic neutron scattering
\cite{24}--\cite{27}, show that the temperature dependence of the
spectrum of elementary excitations $E(p)$, associated with
collective density oscillations
in the $^4$He Bose liquid, is very weak all the way up
to the $\lambda$ point, at all momenta, including the phonon,
maxon, and roton bands. This means that the critical velocity,
determined according to the Landau superfluidity criterion,
$v_c = \min \left[E(p)/p\right]$, hardly changes as $T$ goes up
and does not tend to zero as $T\to T_\lambda$. At the same
time, the breakdown of superfluidity in macroscopic He~II flows
is known \cite{23} to be implied by the processes of creation
of Onsager-Feynman quantum vortices or Anderson closed vortex
rings. As a result, the threshold velocity $v^*_c$ of breakdown of
dissipationless flow as observed in He~II, may be two orders
of magnitude less than the critical velocity
$v_c\simeq \left[\Delta_r/p_r\right]\simeq 60 \mbox{ m/s}$
associated with the roton gap $\Delta_r\simeq 8.6 \mbox{ K}$
in the quasiparticle spectrum $E(p)$ at the point
$p=p_r\simeq 1.9 \mbox{ \AA}^{-1}$.

However, under the conditions where creation and motion of vortices
(or vortex rings) is hindered, much higher values of
threshold velocity can be achieved. For example, in ultrathin
films and capillaries at  $T<1$~K, maximal values of
$v_c^*\simeq (2-3)\mbox{ m/s}$ were observed \cite{23}, and
for the passage of He~II through narrow apertures in thin
partitions, critical velocities $v_c^*\simeq (8-10)\mbox{ m/s}$
were registered \cite{28}, \cite{29}. Moreover, in experiments
on acceleration of ions in He~II \cite{30} at pressures
$P\simeq (15-20)$~bar, threshold velocities of more than
$50 \mbox{ m/s}$ were achieved, close to the roton limit
$v_c\simeq \Delta_r/p_r$. In this connection, the problem of
ab initio theoretical description of the roton minimum
in the quasiparticle spectrum $E(p)$ in liquid helium,
remains topical.

For the first time, such a problem was considered in Ref.~\cite{33}
(see also \cite{34}) within the ``hard-spheres'' model, in which
the Fourier component of the regularized pair interaction
for the $S$ scattering is a sign-changing function of momentum
transfer. Substituting such a potential into the Bogolyubov
quasiparticle spectrum of a weakly nonideal Bose gas \cite{BNN}
leads, under certain conditions, to the appearance of a minimum,
analogous to the roton minimum in the empirical spectrum of
liquid $^4$He. A similar problem was considered in Ref.~\cite{Pash}
for a ``semitransparent spheres'' model with a finite potential
jump. However, for a Bose liquid with strong interaction
between particles and a suppressed BEC $(n_0\ll n)$ the
Bogolyubov approximation is not applicable.

The questions discussed in this paper are both those of the
quantum structure of the SF state of a Bose liquid and the
calculation of the spectrum of elementary excitations based on
a specific form of pair interaction between the bosons.

Our approach is based on the microscopic model \cite{17} of
superfluidity of a Bose liquid with a suppressed BEC and an
intensive pair coherent condensate (PCC), which can arise from
a sufficiently strong effective attraction between bosons in some
domains of momentum space (see below) and is analogous to
the Cooper condensate in a Fermi liquid with attraction between
fermions near the Fermi surface \cite{12}. As a small parameter,
one uses the ratio of the BEC density to the total Bose liquid
density $(n_0 / n)\ll 1$, unlike in the Bogolyubov theory \cite{BNN}
for a quasi-ideal Bose gas, in which the small parameter is
the ratio of the number of supracondensate excitations to the
number of density in the intensive BEC, $(n-n_0)/n_0\ll 1$.

Because of this, the SF state within the model at hand can be
described by a ``truncated'' self-consistent system of Dyson-Belyaev
equations for the normal and anomalous single-particle Green
functions $G_{ij}(k, \omega )$ and the self-energy parts
$\tilde\Sigma_{ij} (k,\omega)$ without account for the diagrams
of second and higher orders in the BEC density.
The renormalized field perturbation theory \cite{10a}--\cite{11}
is used; it is built on combined field variables \cite{31}, \cite{32},
which in the long-wave limit $(p\to 0)$ reduce to the hydrodynamic
variables of macroscopic quantum (at $T=0$) or two-liquid (at $T\ne 0$)
hydrodynamics, whereas in the short-wave band they correspond
to the bosonic quasiparticle creation and annihilation operators.
In this case, the SF component $\rho_s$ is a superposition of the
``exhausted'' single-particle BEC and an intensive ``Cooperlike'' PCC
with coinciding phases (signs) of the corresponding order parameters.

The pair interaction between bosons was chosen in the form of a
regularized repulsion potential in the ``hard spheres'' model
\cite{33}, \cite{34}, whose Fourier component $V(p)$ is an
oscillating sign-changing function of momentum transfer $p$ due
to mutual quantum diffraction of particles.
The same oscillating sign-changing behavior is
characteristic of Fourier components of potentials with finite
jumps or inflection points with an infinite derivative.

As a result of renormalization (``screening'') of the initial
interaction $V(p)$ due to multiparticle collective correlations,
which are described by the boson polarization operator  $\Pi (p, \omega)$,
the interaction gets suppressed in the domains of momentum space
where $V(p)>0$ and enhanced where $V(p)<0$. Such a suppression of
repulsion and enhancement of attraction is implied by the negative
sign of the real part of $\Pi (p, \omega)$ on the ``mass shell''
$\omega=E(p)$ for a decayless quasiparticle spectrum.
It is shown that the integral contribution of the domains of
effective attraction in the renormalized sign-changing interaction
$$\tilde V(p)=V(p)\left[1-V(p)\mbox{Re}\,(\Pi (p, E(p))\right]^{-1}$$
can be sufficient for the formation of an intensive bosonic PCC in
momentum space (although not for the formation of bound boson pairs
in real space).

Self-consistent numerical calculations of the boson self-energy and
polarization operator, pair order parameter, and quasiparticle
spectrum at $T=0$, involving an iteration scheme, have allowed us
to find conditions for the theoretical spectrum $E(p)$ to coincide
with the experimentally observed elementary excitation spectrum
in $^4$He. At the same time it is shown that the roton minimum
in the quasiparticle spectrum $E(p)$ of a Bose liquid is
directly associated with the first negative minimum of the Fourier
component of the renormalized potential of pair interaction between
the bosons (akin to the minimum in the Bogolyubov spectrum \cite{BNN}
of a weakly nonideal dilute Bose gas \cite{33}--\cite{17}).

Finally, the question of applicability of the
Landau criterion to the description of the SF state of
$^4$He in the absence of quantum vortices is discussed.

\section{Green functions and equations for the self-energy parts
in the model of a Bose liquid with a suppressed BEC in the
renormalized perturbation theory}

The main difficulty of the microscopic description of the SF state
of a Bose liquid with a nonzero BEC is the fact that applying
the renormalized perturbation theory directly \cite{7} leads, as
was shown in \cite{10a}-- \cite{11}, to a whole number of divergences
at small energies $\epsilon \to 0$ and momenta $\mathbf{k} \to 0$ and, as
a consequence, to erroneous results in the calculations of different
physical quantities.

Thus, for example, for a Bose system with weak interaction, when the
ratio of the mean potential energy $V(k_0)k_0^3$ ($k_0$ being a
typical momentum transfer) to the corresponding kinetic energy
$k_0^2/2m$ of the bosons is small, the zeroth-approximation polarization
operator $\Pi (\mathbf{k}, \omega)$ and the density-density
response function $\tilde \Pi (\mathbf{k}, \omega)$ calculated to
the first order in the small parameter of interaction $\xi =m k_0V(k_0)\ll1$,
are logarithmically divergent at $k\to 0$, $ \omega\to 0$, whereas
the exact values $\Pi(0,0)$ and $\tilde \Pi(0, 0)$ are finite \cite{10}:
\begin{equation}
\Pi(0, 0)=-\frac{\partial n}{\partial \mu}=-\frac{n}{mc^2}; \qquad
\tilde\Pi(0, 0)=\frac{n}{m}(c_B^2-c^2)\;,
\label{01}
\end{equation}
where $n$ is the total concentration of bosons, $\mu$ the chemical
potential, $c_B=\sqrt{nV_0/m}$ the velocity of sound in the Bogolyubov
approximation for a weakly nonideal Bose gas \cite{BNN},
$V_0\equiv V(0)$ the zeroth Fourier component of the potential,
and $c$ the speed of sound in the $\mathbf{k}\to 0$ limit for the
spectrum of elementary excitations $\epsilon (k)\simeq |\mathbf{k}|c$
in the Belyaev theory \cite{7}:
\begin{equation}
c=\sqrt{\Sigma_{12}(0)/m^*}\;.
\label{02}
\end{equation}
Here $\Sigma_{12}(0)$ is the anomalous self-energy part of bosons
at zero 4-momentum $p\equiv (\mathbf{k}, \epsilon )= 0$,
and $m^*$ is the effective mass of quasiparticles, which is
determined by the relation \cite{8}
\begin{equation}
\frac{1}{m^*}=\frac{2}{B}\left[
\frac{1}{2m}+\frac{\partial \Sigma_{11}(0)}{\partial |\mathbf{k}|^2 }
-\frac{\partial \Sigma_{12}(0)}{\partial |\mathbf{k}|^2 }
\right]\;,
\label{03}
\end{equation}
where $\Sigma_{11}(0)$ and $ \Sigma_{12}(0)$ are, respectively,
the normal and anomalous self-energy parts (at $k\to 0$, $\epsilon \to 0$),
and
\begin{equation}
B=\left[1-\frac{\partial \Sigma_{11}(0)}{\partial\epsilon }\right]^2
-\Sigma_{11}(0)\frac{\partial^2 \Sigma_{12}(0)}{\partial\epsilon^2}+
\frac{1}{2}\frac{\partial^2 }{\partial \epsilon^2}\left[
\Sigma_{12}(0)\right]^2\;.
\label{04}
\end{equation}
The model of a dilute Bose system of hard spheres with a small parameter
$\beta=\sqrt{n/k_0^3}\ll1$, considered in Ref.~\cite{7}, in which,
by means of a summation of the ``ladder'' diagrams with single-particle
Green functions $G_0(p)$ in the zeroth approximation in $\beta$,
it is possible to exclude the infinite repulsion, leads to a finite
value of $\Sigma_{12}(0)$,
\begin{equation}
\Sigma_{12}(0)= \frac{4\pi a_0}{m}n_0\;,
\label{05}
\end{equation}
$a_0$ being the vacuum scattering amplitude of the particles
and $n_0$ the concentration of bosons in the BEC ($\rho_0=mn_0$).

At the same time, in Ref.~\cite{GN}, taking into account an exact
thermodynamic equation
\begin{equation}
\frac{\partial \Sigma_{11}(0)}{\partial \epsilon}=
-\left(
\frac{\partial n_1}{\partial n_0}\right)_{\mu}=
1-\frac{1}{n_0}\Sigma_{12}(0)\frac{d n_0}{d\mu}\;,
\label{07}
\end{equation}
where $ n_1=n-n_0$ is the concentration of supracondensate bosons,
exact asymptotic relations
\begin{equation}
G_{11}(\mathbf{p}\to 0)=-G_{12}(\mathbf{p}\to 0)=
\frac{n_0 mc^2}{n(\epsilon^2-c^2\mathbf{k}^2+i\delta)}\;;
\quad c^2=\frac{n}{m}\frac{d\mu}{dn}
\label{08}
\end{equation}
were obtained.

However, it was shown in Refs.~\cite{10a}, \cite{10} (see also \cite{32})
that at $p\equiv (\mathbf{k},\,\epsilon)=0$ the anomalous self-energy
part is precisely equal to zero, $\Sigma_{12}(0)\equiv 0$.
Problems then emerge with the determination of the velocity of
sound (\ref{02}) and the quasiparticle relaxation time
$\tau\sim \Sigma_{12}^{-1}(0)$ \cite{300}, as well as with the
asymptotic formulas for the normal $G_{11}(p)$ and anomalous
$G_{12}(p)$ Green functions at $p\to 0$ \cite{8}:
\begin{equation}
G_{11}( p\to 0)=-G_{12}(p\to 0)=\frac{\Sigma_{12}(0)}
{B(\epsilon^2-c^2\mathbf{k}^2+i\delta)}\;,
\label{06}
\end{equation}
because at $\Sigma_{12}(0)=0$ the relations (\ref{07}) and (\ref{04})
reduce to the identities
\begin{equation}
\frac{\partial \Sigma_{11}(0)}{\partial \epsilon}\equiv 1\;,
\qquad B\equiv 0\;,
\label{09}
\end{equation}
so that Eqs.~(\ref{06}) and (\ref{02}) with account for (\ref{03})
contain uncertainties of the $0/0$ type.

With the purpose of fixing these controversies, as well as the
infrared divergences and nonanalyticities at $ p\to 0$ emerging in
the nonrenormalized theory, a renormalization procedure for the
field perturbation theory was worked out in Ref.~\cite{11},
employing the method of ``combined variables'' \cite{31}, which
in the long-wave domain $(|\mathbf{k}|<k_0)$ are in fact the
hydrodynamic variables in the spirit of Landau quantum
hydrodynamics \cite{LLD}, while in the short-wave domain
$(|\mathbf{k}|>k_0)$ they reduce to the bosonic creation and annihilation
operators. As was shown in Refs.~\cite{11},\cite{32}, the perturbation
theory built on such ``adequate'' field variables, does not suffer
from infrared divergences at $ p\to 0$, whose source at $T=0$
is the divergence of long-wave quantum fluctuations (acoustic
Goldstone oscillations), associated with a spontaneous breakdown
of continuous gauge and translational symmetries in the SF state
of a Bose system with a uniform coherent condensate. Such oscillations
are essentially the hydrodynamic first sound in liquid $^4$He,
propagating with the velocity of
$c_1\simeq 236 \mbox{ m/s}$.

The choice of combined variables \cite{31}, \cite{32} leads to the
renormalized anomalous self-energy part $\tilde \Sigma_{12}(p)$ which
does not vanish at $ p=0$. Then one can formally restore all the results
of the renormalized field theory \cite{7},\cite{8}, but this time in
terms of the renormalized quantities $\tilde G_{ik}(p)$ and
$\tilde \Sigma_{ik}(p)$, which do not contain singularities at
$p\to 0$ (save for the pole part
$\tilde G_{ik}(p) \sim\vert  p\vert ^{-2}$).
In particular, the square of velocity of first sound $c_1$,
in accordance with Ref.~(\ref{02}), at $T \to 0$ must be equal to
\begin{equation}
c_1^2\equiv \left( \frac{\partial P}{\partial \rho}\right)_\sigma=
\frac{\tilde \Sigma_{12}(0)}{\tilde m^*}\;,
\label{012}
\end{equation}
where the derivative of the pressure $P$ with respect to the
total density $\rho$ is taken at constant entropy $\sigma$,
and the renormalized effective mass $\tilde m^*$ is determined
by the relations (\ref{03}) and (\ref{04}) with
$\tilde \Sigma_{ik}(0)$ substituted for $ \Sigma_{ik}(0)$.

In view of the aforesaid, we will work with the combined variables
\cite{31}, \cite{32},
\begin{equation}
\tilde \Psi(x)= \tilde \Psi_\mathrm{L} (x)+\tilde \Psi_\mathrm{sh} (x)\;,
\label{1}
\end{equation}
where
\begin{equation}
\begin{array}{c}
\displaystyle
\tilde \Psi_\mathrm{L} (x)= \sqrt{\left\langle
\tilde n_\mathrm{L}\right\rangle}\left[
1+\frac{\tilde n_\mathrm{L}-\left\langle \tilde n_\mathrm{L}\right\rangle}
{2\left\langle \tilde n_\mathrm{L}\right\rangle} +i\tilde \phi_\mathrm{L}
\right]; \quad \tilde \Psi_\mathrm{sh}=
\psi_\mathrm{sh}e^{-i\tilde\phi_\mathrm{L}}\;;\\[12pt]
\displaystyle
\psi_\mathrm{sh}=\psi-\psi_\mathrm{L};\quad
\psi_\mathrm{L}(\mathbf{r})=\frac{1}{\sqrt{V}}
\sum_{|\mathbf{k}|<k_0} a_{\mathbf{k}} e^{i\mathbf{k} \mathbf{r}}=
\sqrt{\left\langle \tilde n_\mathrm{L}\right\rangle}
e^{i\tilde \phi_\mathrm{L}}\;.
\end{array}
\label{2}
\end{equation}

Such an approach means that the sepaartion of the Bose system into a
macroscopic coherent condensate and a gas of supracondensate excitations
is made not on the statistical level,
like in the case of a weakly nonideal Bose
gas \cite{BNN}, \cite{7}, but on the level of ab initio field operators,
which are used to construct a microscopic theory of the
Bose liquid. Note that the approximate expression for the long-wave
part $\tilde \Psi_\mathrm{L}$ of the boson field operator $\tilde \Psi$
is given with account only for first-order terms in the expansions
over the slowly changing (hydrodynamic) phase $\tilde \phi_\mathrm{L}$
and the small deviation of the density $\tilde n_\mathrm{L}$ from its
mean value $\left\langle \tilde n_\mathrm{L}\right\rangle$.
In Refs.~\cite{31}, \cite{32} it was assumed that at low $T$,
because of rather weak interaction ($m k_0V(k_0)\ll 1$), almost all
the particles are in the Bose condensate, and therefore the value of
momentum $k_0$ in Ref.~\cite{32} was chosen in such a way that
the approximate equation
$\left\langle \tilde n_\mathrm{L}\right\rangle\simeq n_0$
($n_0$ being the concentration of particles in the BEC) take place.
However, in a Bose liquid with strong interaction, when the
single-particle BEC is strongly suppressed ($n_0\ll n$),
the value $\left\langle \tilde n_\mathrm{L}\right\rangle$ should be normalized
to the density $n_s=\rho_s/m$ of the SF component.

The system of Dyson-Belyaev equations \cite{7},\cite{8}, which allows
one to express the normal $\tilde G_{11}$ and anomalous $\tilde G_{12}$
renormalized single-particle boson Green functions in terms of the
respective self-energy parts $\tilde \Sigma_{11}$ and $\tilde \Sigma_{12}$,
has the form
\begin{equation}
\tilde G_{11} (\mathbf{p},\epsilon)=\left[G_0^{-1}(-\mathbf{p},-\epsilon)-
\tilde \Sigma_{11}(-\mathbf{p},-\epsilon)\right]/Z(\mathbf{p},\epsilon)\;;
\label{3}
\end{equation}
\begin{equation}
\tilde G_{12}(\mathbf{p},\epsilon)=
\tilde \Sigma_{12}(\mathbf{p},\epsilon)/Z(\mathbf{p},\epsilon)\;.
\label{4}
\end{equation}
Here
\begin{equation}
Z(\mathbf{p},\epsilon)=
\left[G_0^{-1}(-\mathbf{p},-\epsilon)-
\tilde \Sigma_{11}(-\mathbf{p},-\epsilon)\right]
\left[G_0^{-1}(\mathbf{p},\epsilon)-
\tilde \Sigma_{11}(\mathbf{p},\epsilon)\right]
-\vert \tilde \Sigma_{12}(\mathbf{p},\epsilon) \vert ^2\;;
\label{5}
\end{equation}
\begin{equation}
G_0^{-1}(\mathbf{p},\epsilon)=
\left[\epsilon-\frac{\mathbf{p}^2}{2m}+\mu+i\delta \right];\qquad
(\delta\to +0)\;,
\label{6}
\end{equation}
where $\mu$ is the chemical potential of the quasiparticles,
which satisfies the Hugengoltz-Peins relation \cite{19}:
\begin{equation}
\mu=\tilde\Sigma_{11}(0,0)-\tilde\Sigma_{12}(0,0)\;.
\label{7}
\end{equation}

Due to a strong hybridization of the single-particle and collective
branches of elementary excitations in the Bose liquid with a
finite BEC $ (n_0\ne 0)$, the poles of the two-particle and all
many-particle Green functions, as well as the full four-pole
function of the pair interaction of bosons
$ \tilde \Gamma ( p_1,  p_2, \mathop{ p_3}, \mathop{p_4})$,
coincide with the poles of the single-particle Green functions
$ \tilde G_{ik}(\mathbf{p}, \epsilon)$
\cite{8}, \cite{FF}.

Therefore the spectrum of all elementary excitations with zero
spirality is determined by the zeros of the function
$Z(\mathbf{p}, \epsilon )$:
\begin{equation}
E(\mathbf{p})=\left\{ \left[\frac{\mathbf{p}^2}{2m}+\tilde \Sigma_{11}^s
(\mathbf{p}, E(\mathbf{p})) -\mu \right]^2- \vert \tilde \Sigma _{12}
(\mathbf{p},  E(\mathbf{p})) \vert ^2\right\}^{1/2}
+\tilde \Sigma_{11}^a (\mathbf{p}, E(\mathbf{p}))\;,
\label{8}
\end{equation}
where
$$
\tilde \Sigma_{11}^{s,a} (\mathbf{p}, \epsilon) =\frac{1}{2}
\left[\tilde \Sigma_{11}(\mathbf{p}, \epsilon) \pm
\tilde \Sigma_{11}(-\mathbf{p}, -\epsilon) \right]\;,$$
the $(+)$ sign corresponding to the symmetric part of $\tilde \Sigma_{11}^s$,
the $(-)$ sign---to the antisymmetric part
$\tilde \Sigma_{11}^a$. In the sequel, we will assume that
$\tilde \Sigma_{11}$ is an even function of $\mathbf{p}$ and $\epsilon$,
so that $\tilde \Sigma_{11}^a=0$ and
$\tilde \Sigma_{11}^s=\tilde \Sigma_{11}$.
Then, relation (\ref{7}) ensures the acoustic dispersion law for the
quasiparticles at $\mathbf{p}\to 0$, and the first derivative of
$\tilde \Sigma_{11}$ over $\epsilon$ in Eq.~(\ref{04}) at $\epsilon \to 0$
and $ p \to 0$ vanishes:
$\frac{\partial \tilde \Sigma_{11}(0,0)}{\partial \epsilon}=0$, so that
$B\ne 0$, while $\tilde \Sigma_{12}(0)=n_0\frac{d\mu}{d n_0}$, in
accordance with Eq.~(\ref{06}).

As was shown in Ref.~\cite{17}, for a Bose liquid with strong enough
interaction between particles, when the BEC is strongly suppressed,
one can, when defining $\tilde \Sigma_{ik}(\mathbf{p}, \epsilon)$ in the
form of a sequence of irreducible diagrams containing condensate
lines \cite{7}, restrict oneself, with good precision, to the
first (lowest) terms in the expansion over the small BEC density
($n_0\ll n$). Such an approximation is exactly opposite to the
Bogolyubov approximation \cite{BNN} for a weakly nonideal Bose gas with
an intensive BEC, when $n_0\simeq n$.

As a result, up to terms of first order in the small parameter
$n_0/n\ll 1$, for a Bose liquid one gets the ``trimmed''
system of equations for $\tilde{\Sigma}_{ik}$ \cite{17}:
\begin{equation}
\tilde \Sigma_{11}(\mathbf{p}, \epsilon) =n_0\Lambda (\mathbf{p}, \epsilon)
 \tilde V(\mathbf{p}, \epsilon)+
 n_1 V(0)+\tilde \Psi _{11}(\mathbf{p}, \epsilon)\;;
\label{15}
\end{equation}
\begin{equation}
\tilde \Sigma_{12}(\mathbf{p}, \epsilon) =
n_0\Lambda (\mathbf{p}, \epsilon) \tilde V(\mathbf{p}, \epsilon)+
\tilde \Psi_{12} (\mathbf{p}, \epsilon)\;,
\label{16}
\end{equation}
where
\begin{equation}
\tilde{\Psi}_{ij}(\mathbf{p}, \epsilon)=
i\int \frac{d^3 \mathbf{k}}{(2\pi)^3}\int \frac{d\omega}{2\pi}
G_{ij}(\mathbf{k})\tilde V (\mathbf{p}-\mathbf{k}, \epsilon-\omega)
\Gamma(\mathbf{p},\epsilon, \mathbf{k}, \omega )\;,
\label{17}
\end{equation}
\begin{equation}
\tilde V(\mathbf{p}, \epsilon) =V(\mathbf{p})\left[1- V(\mathbf{p})
 \Pi(\mathbf{p}, \epsilon)\right]^{-1}\;.
\label{18}
\end{equation}

Here $V(\mathbf{p})$ is the Fourier component
of the input potential of pair interaction of bosons,
$\tilde V(\mathbf{p}, \epsilon)$ is the renormalized
(``screened'' due to multiparticle collective effects)
Fourier component of the retarded (nonlocal) interaction;
$\Pi (\mathbf{p}, \epsilon)$ is the boson polarization operator:
\begin{equation}
\begin{array}{c}
\displaystyle
\Pi(\mathbf{p}, \epsilon)=
i\int \frac{d^3 \mathbf{k}}{(2\pi)^3}\int \frac{d\omega}{2\pi}
\Gamma(\mathbf{p}, \epsilon,\mathbf{k},\omega)
\\[12pt]
{}\times\left\{G_{11}(\mathbf{k}, \omega)
G_{11}(\mathbf{k}+\mathbf{p}, \epsilon+\omega)
+G_{12}(\mathbf{k}, \omega)
G_{12}(\mathbf{k}+\mathbf{p}, \epsilon+\omega)\right\}\;;
\end{array}
\label{19}
\end{equation}
$\Gamma (\mathbf{p},\, \epsilon ;\,\mathbf{k},\, \omega)$ is the vertex
part (three-pole function), which describes multiparticle correlations;
$\Lambda (\mathbf{p}, \epsilon)=\Gamma (\mathbf{p}, \epsilon, 0, 0)=
\Gamma (0,0,\mathbf{p}, \epsilon)$ ,\,
and $n_1$ is the number of supracondensate particles
($n_1\gg n_0$), which is determined from the condition of
conservation of the total number of particles:
\begin{equation}
n=n_0+n_1=n_0+ i\int\frac{d^3 \mathbf{k}}{(2\pi)^3}\int \frac{d\omega}{2\pi}
G_{11}(\mathbf{k}, \omega)\;.
\label{19-1}
\end{equation}
In the sequel, as well as in Ref.~\cite{17},
in the integral relations (\ref{17}) and (\ref{19})
we will only take into account the residues at the
poles of single-particle Green functions
$\tilde G_{ij}(\mathbf{p}, \epsilon)$,
neglecting the contributions of eventual poles of the functions
$\Gamma (\mathbf{p},\epsilon, \mathbf{k}, \omega )$
and $\tilde V(\mathbf{p}, \epsilon)$,
which do not coincide with the poles of
$\tilde G_{ij}(\mathbf{p}, \epsilon )$. As a result,
taking into account relations (\ref{3})--(\ref{6}),
(\ref{8}), (\ref{15}) and (\ref{18}), Eqs.~(\ref{17}) on the mass shell
$\epsilon=E(\mathbf{p})$ assume the following form (at $T=0$):
\begin{equation}
\begin{array}{c}
\displaystyle
 \tilde{\Psi}_{11}(p)\equiv\tilde{\Psi}_{11}(\mathbf{p}, E(p))=
\frac{1}{2}\int \frac{d^3 \mathbf{k}}{(2\pi)^3} \Gamma (\mathbf{p}, E(\mathbf{p});
\mathbf{k}, E(\mathbf{k}))
 \\[12pt]
\displaystyle
{} \times \tilde{V}(\mathbf{p}- \mathbf{k},
E (\mathbf{p})- E(\mathbf{k})) \left[\frac{A(\mathbf{k})}
{E(\mathbf{k})}-1\right]\;;
\end{array}
\label{20}
\end{equation}
\begin{equation}
\begin{array}{c}
\displaystyle \tilde{\Psi}_{12}(p)\equiv
\tilde{\Psi}_{12}(\mathbf{p},  E(p))=
 -\frac{1}{2}\int \frac{d^3  \mathbf{k}}
{(2\pi)^3} \Gamma (\mathbf{p}, E(\mathbf{p});
\mathbf{k}, E(\mathbf{k})) \tilde{V}(\mathbf{p}- \mathbf{k},
E(\mathbf{p})- E (\mathbf{k}))\\[12pt]
\displaystyle
{}\times \frac{n_0\Lambda (\mathbf{k}, E(\mathbf{k}))
\tilde{V}(\mathbf{k}, E(\mathbf{k}))+
\tilde{\Psi}_{12}(\mathbf{k})}{E(\mathbf{k})}\;,
\end{array}
\label{21}
\end{equation}
where
\begin{equation}
A(\mathbf{p})=n_0\Lambda (\mathbf{p}, E(\mathbf{p}))
\tilde{V}(\mathbf{p}, E(\mathbf{p}))+
n_1V(0) +\tilde{\Psi}_{11}(\mathbf{p}) +
\frac{\mathbf{p}^2}{2m}-\mu\;.
\label{23}
\end{equation}
Then the nonlinear equation (\ref{8}) for the quasiparticle spectrum
$E(\mathbf{p})$, according to Eqs.~(\ref{15}), (\ref{16}), takes the form
\begin{equation}
E(\mathbf{p})=\sqrt{A^2(\mathbf{p})-\left[
n_0\Lambda (\mathbf{p}, E(\mathbf{p})) \tilde{V}(\mathbf{p}, E(\mathbf{p}))+
\tilde{\Psi}_{12}(\mathbf{p})\right]^2}\;,
\label{22}
\end{equation}
and the total quasiparticle concentration in the Bose liquid is
determined by the relation
\begin{equation}
n=n_0+\frac{1}{2} \int \frac{d^3 \mathbf{k}}{(2\pi)^3}
\left[\frac{A(\mathbf{k})}{E(\mathbf{k})}-1\right]\;.
\label{25a}
\end{equation}

The Hugengoltz-Peins relation (\ref{7}),
according to Eqs.~(\ref{15}) and (\ref{16}), can be
represented as
\begin{equation}
\mu=n_1 V(0)+\tilde{\Psi}_{11} (0)-\tilde{\Psi}_{12}(0)\;,
\label{25}
\end{equation}
as a result of which Eq.~(\ref{23} takes the form
\begin{equation}
A(\mathbf{p})=n_0\Lambda (\mathbf{p}, E(\mathbf{p}))\tilde{V}
(\mathbf{p}, E(\mathbf{p}))+
\left[\tilde{\Psi}_{11}(\mathbf{p})
-\tilde{\Psi}_{11}(0)\right]
+\tilde{\Psi}_{12}(0)+\frac{\mathbf{p}^2}{2m}\;.
\label{26}
\end{equation}

From Eqs.~(\ref{22}) and (\ref{26}) it follows that the quasiparticle
spectrum, because of the analyticity of the functions
$\tilde \Psi_{ij}(\mathbf{p}, \epsilon)$, is acoustic at $\mathbf{p}\to 0$,
and its structure at $\mathbf{p}\ne 0$ depends essentially on the
character of the renormalized pair interaction of bosons.

If one assumes that the functions
$\Psi_{11}$ and $ \Psi_{12}$ depend weakly
on the explicit form of the quasiparticle spectrum $E(\mathbf{p})$,
then by virtue of Eq.~(\ref{04}), one can, with good precision,
assert $B=1$, so that the effective mass, according to
Eqs.~(\ref{03}), (\ref{15}), (\ref{16}), is determined by the
expression
\begin{equation}
\frac{1}{\tilde m^*}=\frac{1}{m}+\frac{\partial^2 \tilde \Psi_{11}(0) }
{\partial \vert \mathbf{p}\vert ^2}-\frac{\partial^2 \tilde \Psi_{12}(0) }
{\partial \vert \mathbf{p}\vert ^2}
\label{26-1}
\end{equation}
and the expression for the velocity of sound, according to
Eqs.~(\ref{22}) and (\ref{26}), can be cast in the form
\begin{equation}
 c=\sqrt{\Lambda (0,0)\tilde V(0,0) \tilde n/\tilde m^*}\;;
\qquad  \tilde n=n_0+\frac{\tilde \Psi_{12}(0)}
{\Lambda (0,0)\tilde V(0,0)}\;,
\label{28p}
\end{equation}
which is analogous to the expression for the Bogolyubov velocity
of sound for a weakly nonideal Bose gas
$c_{\mbox{\small B}}=\sqrt{V(0) n/m}$.
Then the conditions $\tilde n>0$ and $c=c_1$ imply severe
constraints on the choice of the parameters of the model
of interaction of the bosons (see below).

Indeed, at $\mathbf{p}\to 0$ and $E(\mathbf{p})=c_1 \vert \mathbf{p} \vert \to 0$,
Eq.~(\ref{21}), due to the the momentum
dependence of the spectrum $E(\mathbf{p})$ and the functions
$\tilde V(\mathbf{p})\equiv \tilde V(\mathbf{p}, E(\mathbf{p}))$,
$\Lambda (\mathbf{p})\equiv \Lambda (\mathbf{p}, E(\mathbf{p}))$ and
$\tilde \Psi_{12}(\mathbf{p})$ being isotropic, assumes the form
\begin{equation}
\tilde \Psi_{12}(0)=- \frac{1}{(2\pi )^2}\int\limits_{0}^{\infty}
\frac{ k^2dk}{E(\mathbf{k})}
\left[ n_0\Lambda^2(\mathbf{k})\tilde V^2(\mathbf{k})+\Lambda (\mathbf{k})
\tilde V (\mathbf{k})
\tilde \Psi_{12}(\mathbf{k})\right]\;.
\label{28-1}
\end{equation}
It follows that the first integral addend on the right-hand side
of Eq.~(\ref{28-1}) is always negative, so that the value
$\tilde \Psi_{12}(0)$, which plays the role of the pair order
parameter \cite{17}, can also be negative
at not too small values of $n_0$, regardless of the sign of
the renormalized interaction $\Lambda(\mathbf{k}) \tilde V(\mathbf{k})$.
The condition $\tilde \Psi_{12}(0)<0$ means that the phase of
the PCC is opposite to that of the BEC, because $n_0>0$.
Moreover, in this case, due to the condition $\Lambda(0) \tilde V(0)>0$,
which ensures that the system is globally stable against a
spontaneous collapse, at sufficiently small densities of the BEC,
in accordance with Ref.~(\ref{28p}), the values $\tilde n$ and
$\tilde c^2$ can become negative, which corresponds to an
instability in the phonon spectrum, which occurs when
$\vert \Psi_{12}(0)\vert >n_0\Lambda (0) V(0)$.

However, if the pair interaction between bosons in a broad enough
region of the momentum space has, due to some reason, the
character of attraction, i.e.,
$\Lambda(\mathbf{k}) \tilde V(\mathbf{k})<0$ at $\mathbf{k}\ne 0$,
and if the magnitude of that attraction is
large enough (see below) and the BEC density is small enough
$(n_0\ll n)$, the second (positive) addend on the right-hand side
of Eq.~(\ref{28-1}) can outweigh the first (negative) one. Then
$\tilde \Psi_{12}(0)$ will be positive, and the phase
of the PCC will coincide with that of the BEC, so that $\tilde n
>0$ and $\tilde c^2>0$.

Note that when the BEC is totally absent $(n_0=0)$, when the
integral equation (\ref{21}) is homogeneous, i.e., degenerate
with respect to the phase $\tilde \Psi_{12}(\mathbf{p})$, the
condition of stability of the phonon spectrum $\tilde c^2=\tilde
\Psi_{12}(0)/m^* >0$ is secured by the choice of the respective
sign (phase) of the pair order parameter $\tilde \Psi_{12}(0)>0$
(see \cite{17}).

On the other hand, since at $T=0$ the density of the SF component
$\rho_s$ coincides with the total density $\rho=mn$ of the Bose
liquid, assuming $\tilde n=n$ and with account for (\ref{16}), one
gets the following relations:
\begin{equation}
\rho_s=\rho_0+\tilde \rho_s=m \frac{\tilde\Sigma_{12}(0)} {\Lambda
(0)\tilde V(0)}\;;
\label{29p}
\end{equation}
\begin{equation}
\tilde \rho_s=m n_1=m \frac{\tilde\Psi_{12}(0)}
{\Lambda (0)\tilde V(0)}\;, \label{30p}
\end{equation}
where $\rho_0=m n_0$ is the single-particle BEC density, and
$\tilde \rho_s$ is the density of the ``Cooper'' PCC. The
concentration $n_1=n-n_0$ is then determined from the relation
(\ref{25a}), and for liquid $^4$He at $T\to 0$, in accordance with
the experimental data \cite{9}, it should be not less than $90\%$
of the full concentration of $^4$He atoms. Thus, the SF component
in this model is a superposition of a single-particle and a pair
coherent condensates, and relations (\ref{25a}) and (\ref{29p})
impose additional constraints on the parameters of the microscopic theory
of the SF Bose liquid.

\section{Influence of a renormalized pair interaction on the
spectrum of elementary excitations in a Bose liquid with a
suppressed BEC}

To describe interaction of helium atoms in real space,
various phenomenological and semi-empirical potentials
\cite{T6}--\cite{enciklop} are conventionally used, which describe
strong repulsion at small distances and weak van der Waals
attraction at large distances (see Table I). As was shown in
\cite{MashVil} by means of a numerical solution of the
Schr\"odinger equation, some of those potentials give rise to
bound states---discrete levels with very small binding energy
($\Delta E <0.1$~K).

However, all these potentials are characterized by a divergence
$(\pm \infty )$ at  $r\to 0$ and are not suitable for the
description of pair interaction in momentum space, since the
respective Fourier components are infinite.
Moreover, in a condensed state one cannot directly use the free
atomic potentials---it is rather necessary to address the problem
of constructing adequate pseudopotentials \cite{psevdo}.
Therefore, as the input potential of pair interaction,
a regularized ``hard-spheres'' potential \cite{33}, \cite{34} with
the radius equal to the quantum radius of the helium atom,
$r_0=1.22$~\AA, will be employed.

As was shown in Refs.~\cite{33}, \cite{34}, for spherically
symmetric scattering ($S$-wave), taking into account that the
radial wave function $u(r)$ of relative motion of particles
vanishes at the infinite jump of the potential $V(r)\to \infty$
at $r= 2.44$~\AA, as well as the boundary condition
$V(r)u(r)=\lambda\delta (r-a)$, the Fourier component of the pair
potential is a finite sign-changing function of momentum transfer:
\begin{equation}
V(p)=V_0j_0(pa)\;; \qquad j_0(x)=\frac{\sin x}{x}\;.
\label{31p}
\end{equation}
Here $j_0(x)$ is the spherical Bessel function of zeroth order,
and $V_0$ is a positive constant which is determined in a self-consistent
way from a nonlinear integral equation for the single-particle
Green function at $\mathbf{p}\to 0$ and depends on the dimensionless
Bose liquid density $na^3$ (see Refs.~\cite{33}, \cite{34}).

The potential (\ref{31p}) is shown with a dashed line in Fig.~1.
It corresponds to repulsion, $V(p)>0$, in those regions of the
momentum space where $\sin (pa)>0$ (in particular, when $pa<\pi$),
and to attraction, $V(p)<0$, in those ones where $\sin(pa)<0$ (for
example, $\pi<pa<2\pi$).

As a different example, consider the Fourier component of a
model potential in the form of a ``Fermi'' function
\begin{equation}
V(r)=V_0\left\{\exp \left( \frac{r^2-a^2}{b^2}\right)+1\right\}^{-1}\;,
\label{F}
\end{equation}
which at $b=0$ degenerates into a ``step'' of a
finite height $V_0$ at $r<a$, corresponding to a model of
semitransparent spheres \cite{Pash}.
In this case the Fourier component is expressed in terms of the
first order spherical Bessel function:
\begin{equation}
V(p)=V_0\frac{j_1(pa)}{pa};\qquad
j_1(x)=\frac{\sin(x)-x \cos(x)}{x^2}\;.
\label{F1}
\end{equation}
The Fourier component of a smooth potential $V(r)$ in the form of
a Lindhardt function \cite{12}, whose derivative at the
inflection point $r=a$ is $-\infty$, has the same shape
(see Appendix A).
The sign-changing oscillations of the Fourier component $V(p)$ in
the momentum space are then formally analogous to the
Ruderman-Kittel-Ksui-Yoshida oscillations \cite{Wite} in exchange
interaction of spins or to the Friedel oscillations \cite{12}
of the screened Coulomb potential with period $\pi/k_F$ in real
space, which arise as a result of scattering of electrons
(fermions) on a Fermi sphere, filled according to the Pauli
principle, of diameter $2k_F$ ($k_F$ being the electron Fermi
momentum).

It should be stressed that the sign-changing Friedel oscillations
are not directly associated with the jump of the Fermi electron
distribution function at $T=0$, but rather with a relatively weak
singularity (inflection point) in the momentum dependence of
the static polarization operator $\Pi_c(\mathbf{p}, 0)$, which is
characterized by a logarithmic divergence at $p=2k_F$ of the first
derivative of $\Pi_c(\mathbf{p}, 0)$ with respect to $p$.
This means that the sign-changing oscillations of the Fourier
component of the potential $V(r)$ can arise not only in the ``hard
spheres'' model with an infinite jump of $V(r)$, but also with
finite jumps of the potential, or for ``smooth'' potentials with
weak fractures or inflection points with an infinite
derivative with respect to $r$.

Thus, the existence of negative values $V(p)<0$, i.e., of an
effective attraction in some regions of the momentum space, is a
consequence of certain peculiarities (jumps, fractures, inflection
points) in the radial dependence of the pair interaction
potentials, and is of quantum mechanical nature.

If one substitutes the oscillating potential (\ref{31p}) (or (\ref{F1}))
into the Bogolyubov spectrum of a dilute quasi-ideal Bose gas \cite{BNN}
\begin{equation}
E_B(\mathbf{p})=\left\{\frac{\mathbf{p}^2}{2m}\left[\frac{\mathbf{p}^2}{2m}+
2nV(p)\right]\right\}^{1/2},
\label{32p}
\end{equation}
then, by choosing two parameters, $V_0$ and $a$,
independently, one can achieve a rather satisfactory coincidence
of the spectrum $E_B(\mathbf{p})$ with the elementary excitation
spectrum $E_\mathrm{exp}(p)$ in liquid $^4$He derived from
neutron scattering experiments (Fig.~2, solid curve).
However, the self-consistent solution for $a=2.2$~\AA{} and $na^3\simeq
0.23$, which was obtained in Refs.~\cite{33}, \cite{34} in the
framework of the regularized ``hard spheres'' model, differs
considerably from $E_\mathrm{exp}(\mathbf{p})$, and for a more realistic
parameter $a=2.44$~\AA, characteristic of $^4$He, when $na^3=0.315$,
the spectrum (\ref{32p}) with the potential (\ref{31p}) turns out
to be unstable, because $E_B^2(\mathbf{p})<0$ in some range of $p$
(Fig.~2, dashed curve), which points out that the Bogolyubov
theory \cite{BNN} is unapplicable for the description of the $^4$He Bose
liquid.

On the other hand, multiparticle collective effects in the Bose
liquid, according to Ref.~(\ref{18}), lead to an essential
renormalization of the pair interaction, which determines the
normal and anomalous self-energy parts, Eqs.~(\ref{15}) and (\ref{16}).
Taking into account Eqs.~(\ref{18}) and (\ref{31p}), the vertex
part (four-point function) of the retarded interaction between
bosons assumes the form
\begin{equation}
\tilde V (\mathbf{p},\omega)
=\frac{V_0 \sin (pa)}{pa-V_0\Pi (\mathbf{p}, \omega) \sin (pa)},
\label{33p}
\end{equation}
where $\Pi (\mathbf{p}, \omega)$ is the bosonic polarization operator
(\ref{19}) in the SF state, which, with account for the pole parts
of the Green functions (\ref{3}) and (\ref{4}), is calculated in
Appendix B.

An important feature of the renormalized interaction (\ref{33p})
is that in those regions of phase volume $(\mathbf{p}, \omega)$ in
which $\Pi (\mathbf{p}, \omega)<0$, the repulsion (when $\sin (pa)>0$)
gets suppressed while the attraction (when $\sin (pa)<0$) gets
effectively enhanced.

As follows from Eqs.~(\ref{22}) and (\ref{26}), the main influence
on the quasiparticle spectrum $E(\mathbf{p})$ comes from the shape of
the interaction potential (\ref{33p}) on the ``mass shell'', when
$\omega=E(\mathbf{p})$. The real part of the polarization operator
then has the form (see Appendix B):
\begin{equation}
\begin{array}{c}
\displaystyle
\mbox{Re} \,\Pi(\mathbf{p}, E(\mathbf{p}))=\frac{1}{2}
\int\frac{d^3\mathbf{k}}{(2\pi)^3}
\frac{1}{E(\mathbf{k})-E(\mathbf{k}-\mathbf{p})-E(\mathbf{p})} \\[12pt]
\displaystyle
{}\times\left\{\frac{F_{-}(\mathbf{k}, \mathbf{p})}
{E(\mathbf{k})[E(\mathbf{k})+E(\mathbf{k}-\mathbf{p})-E(\mathbf{p})]} -
\frac{F_{+}(\mathbf{k}, \mathbf{p})}
{E(\mathbf{k})[E(\mathbf{k})+
E(\mathbf{k}-\mathbf{p})+E(\mathbf{p})]}\right\}
\label{PI1}
\end{array}
\end{equation}
where
\begin{equation}
\begin{array}{c}
\displaystyle
F_{-}(\mathbf{k}, \mathbf{p})=\left[
E(\mathbf{k})+\frac{\mathbf{k}^2}{2m}-\mu+
\tilde \Sigma_{11}(\mathbf{k}, E(\mathbf{k}))\right]
\\[12pt]
\displaystyle{}\times\left[
E(\mathbf{k})-E(\mathbf{p})+\frac{(\mathbf{k}-\mathbf{p})^2}{2m}-\mu+
\tilde \Sigma_{11}(\mathbf{k}-\mathbf{p},
E(\mathbf{k})-E(\mathbf{p}))\right]\\[12pt]
{}+\tilde \Sigma_{12}(\mathbf{k}, E(\mathbf{k}))
\tilde \Sigma_{12}(\mathbf{k}-\mathbf{p}, E(\mathbf{k})-E(\mathbf{p}))\;,
\label{PI2}
\end{array}
\end{equation}

\begin{equation}
\begin{array}{c}
\displaystyle
F_{+}(\mathbf{k}, \mathbf{p})=\left[
E(\mathbf{k}-\mathbf{p})+\frac{(\mathbf{k}-\mathbf{p})^2}{2m}-\mu+
\tilde \Sigma_{11}(\mathbf{k}-\mathbf{p}, E(\mathbf{k}-\mathbf{p}))\right]
\\[12pt]
\displaystyle
{}\times\left[
E(\mathbf{k}-\mathbf{p})+E(\mathbf{p})+\frac{\mathbf{k}^2}{2m}-\mu+
\tilde \Sigma_{11}(\mathbf{k},
E(\mathbf{k}-\mathbf{p})+E(\mathbf{p}))\right]\\[12pt]
{}+\tilde \Sigma_{12}(\mathbf{k}-\mathbf{p}, E(\mathbf{k}-\mathbf{p}))
\tilde \Sigma_{12}(\mathbf{k}, E(\mathbf{k}-\mathbf{p})+E(\mathbf{p}))\;.
\label{PI3}
\end{array}
\end{equation}

The key feature of the screened potential
$\tilde V(\mathbf{p}, E(\mathbf{p}))$
is the fact that $\mbox{Re}\, (\Pi(\mathbf{p}, E(p))<0$
for all $\mathbf{p}>0$, if the quasiparticle spectrum $E(\mathbf{p})$ is
stable with respect to decays into a pair of quasiparticles \cite{LP},
\cite{8}, i.e., if for all $\mathbf{p}$ and $\mathbf{k}$ the conditions
\begin{equation}
E(\mathbf{p})<E(\mathbf{k})+E(\mathbf{k}-\mathbf{p})\;; \qquad
E(\mathbf{k})<E(\mathbf{p})+E(\mathbf{k}-\mathbf{p})
\label{34p}
\end{equation}
are fulfilled.

Indeed, as follows from Eq.~(\ref{PI1}), the common denominator in
front of the curly braces is always negative,
\begin{equation}
\left[E(\mathbf{k})-E(\mathbf{k}-\mathbf{p})-E(\mathbf{p})\right]<0\;,
\label{z1}
\end{equation}
whereas the denominator in the first term in the curly braces is
always positive,
\begin{equation}
\left[E(\mathbf{k})+E(\mathbf{k}-\mathbf{p})-E(\mathbf{p})\right]>0
\label{z2}
\end{equation}
and smaller than the positive denominator in the second term
\begin{equation}
\left[E(\mathbf{k})+E(\mathbf{k}-\mathbf{p})+E(\mathbf{p})\right]>0\;.
\label{z3}
\end{equation}

At the same time, as numerical simulation has shown,
the numerators $F_{-}$ and $F_{+}$ in both terms remain positive
for any $\mathbf{p}$ and $\mathbf{k}$ (see Fig.~3a).
Therefore, the common sign of integrands in Eq.~(\ref{PI1})
is negative, so that $\mbox{Re}\,(\Pi(\mathbf{p}, E(p))<0$.
Figure~3b shows the $\mathbf{p}$ dependence
of $\mbox{Re}\,(\Pi(\mathbf{p}, E(\mathbf{p}))$, obtained
numerically by iterations.

Taking into account the negative sign and the relatively weak
momentum dependence of $\mbox{Re}\,(\Pi(\mathbf{p}, E(\mathbf{p}))$
in a wide region $p\ne 0$, one can approximate the renormalized potential
(\ref{33p}) at $\omega=E(\mathbf{p})$ in the integrands of Eqs.~(\ref{20}),
(\ref{21})
with a more simple one:
\begin{equation}
\tilde V (p)=\frac{V_0 \sin (pa)}{pa+\alpha \sin (pa)},
\label{35p}
\end{equation}
where $\alpha=V_0 \vert\tilde \Pi\vert$, and $\vert\tilde \Pi\vert=
\overline{\vert \mbox{Re}\,(\Pi(\mathbf{p},E(\mathbf{p})) \vert}$ 
is the mean absolute value of the real part of the polarization
operator on the mass shell in the domain of existence of the
spectrum $E(\mathbf{p})$. In what follows, all the numerical
simulations are based on the model potential (\ref{35p}),
in which the quantities $V_0$ and $\alpha$ are free fitting
parameters. Figure 1 depicts the function $V(p)/V_0$ for
different values of the dimensionless parameter $\alpha$.
As far as the vertices $\Lambda$ and $\Gamma$ are concerned,
their relatively weak $\mathbf{p}$ and $\omega$ dependence may be
disregarded, putting
$\Lambda\simeq \Gamma \simeq \Lambda(0,0)=\mbox{const}$ and
incorporating the constant value of $\Lambda(0,0)$ into $V_0$.  

As a result, equations (\ref{20}) and (\ref{21}) for the
functions $\tilde\Psi_{ij}$ boil down to a simpler form,
\begin{equation}
\tilde\Psi_{11}(\mathbf{p})=\frac{1}{2}\int\frac{d^3 \mathbf{k}}{(2\pi)^3}
\tilde V(\mathbf{p}-\mathbf{k})\left[\frac{A_0(\mathbf{k})}{E(\mathbf{k})}-1\right]\;,
\label{36p}
\end{equation}
\begin{equation}
\tilde\Psi_{12}(p)=-\frac{1}{2}\int\frac{d^3 \mathbf{k}}{(2\pi)^3}
\tilde V(\mathbf{p}-\mathbf{k})\frac{n_0\tilde V(\mathbf{k})+\Psi_{12}(\mathbf{k})}
{E(\mathbf{k})}\;,
\label{37p}
\end{equation}
where
\begin{equation}
E(\mathbf{p})=\sqrt{A_0^2(\mathbf{p})-
\left[n_0\tilde V(\mathbf{p})+\Psi_{12}(\mathbf{p})\right]^2}\;;
\label{38p}
\end{equation}
\begin{equation}
A_0(\mathbf{p})=n_0\tilde V(\mathbf{p})+\left[\tilde\Psi_{11}(\mathbf{p})-
\tilde\Psi_{11}(0)\right]+\tilde\Psi_{12}(0)+\frac{\mathbf{p}^2}{2m}\;.
\label{39p}
\end{equation}
In this case, for $\mathbf{p}\to 0$, from Eq.~(\ref{38p}), taking
into account (\ref{39p}) and (\ref{26-1}), one gets
\begin{equation}
E(\mathbf{p}\to 0)=\vert \mathbf{p}\vert
\sqrt{\left[n_0\tilde V(0)+\tilde\Psi_{12}(0)\right]/\tilde m^*}\;,
\label{40p}
\end{equation}
where $\tilde V(0)=V_0/[1-V_0\Pi(0,0)]$, and $\Pi(0,0)$ is determined
by the first relation in (\ref{01}),
so that for the velocity of sound one has the following relation
(see~(\ref{28p})):
\begin{equation}
c^2=\frac{1}{\tilde m^*}\left[
\frac{n_0 V_0}{1+n_0V_0/mc^2}+\tilde \Psi_{12}(0)\right]\;,
\label{sq}
\end{equation}
which is equivalent to a biquadratic equation for $c$. 
Under the conditions $nV_0\gg mc^2$,  $n_0\ll n$, from Eq.~(\ref{sq})
there follows an approximate expression 
\begin{equation}
c^2\simeq \frac{\tilde \Psi_{12}(0)}{\tilde m^*}
\left( 1+\frac{m n_0}{\tilde m^* n}\right) \;.
\label{sq1}
\end{equation}
The parameters $V_0$ and $\alpha$ were chosen in such a way
that the phase velocity $E(\vert \mathbf{p}\vert\to 0)/\vert \mathbf{p}\vert$
coincide with the hydrodynamic sound velocity $c_1\simeq 236$~m/s
in liquid $^4$He.
On the other hand, the choice of those parameters was to ensure
the maximal coincidence of the spectrum $E(\mathbf{p})$ with
the experimental spectrum $E_\mathrm{exp}(\mathbf{p})$ in $^4$He
\cite{24}--\cite{27} at all values of $p$.

Figure 4 (a,b,c) depicts the momentum dependence of the functions
$\tilde\Psi_{11}(\mathbf{p})$, $\tilde\Psi_{12}(\mathbf{p})$ and
$A_0(p)$, calculated according to Eqs.~(\ref{36p}), (\ref{37p})
and (\ref{39p}) for the values of
$\alpha=4.4$, $V_0/(4\pi^2 a^3)=17.7$~K  at $a=2.44$~\AA,
and Fig.~5a presents the quasiparticle spectrum $E(p)$ obtained
from  Eq.~(\ref{38p}) for those same values of the parameters.
It can be seen that this spectrum is in qualitative agreement
with the experimental spectrum $E_\mathrm{exp}$  in $^4$He,
however the numerical correspondence of the positions and values
of the maximum and minimum of quasiparticle energy,
$E_\mathrm{max}=15.9$~K at $p_\mathrm{max}=1.05 \mbox{ \AA}^{-1}$
and $E_\mathrm{min}=10.7$~K at $p_\mathrm{min}=1.72 \mbox{ \AA}^{-1}$,
cannot be deemed satisfactory. Besides, the sound velocity calculated
according to Eqs.~(\ref{28p}) and (\ref{sq}) turns out to be too low,
$c=2.08 \times 10^4 \mbox{ cm/s}$, while the total particle concentration
calculated according to the formula (\ref{25a}) is too high,
$n=2.57 \times 10^{22} \mbox{ cm}^{-3}$, the concentration of the particles
in the BEC being low, $n_0=0.03n$.

There is a much better numerical agreement at the values of
$\alpha=4.52$,  $V_0/(4\pi^2 a^3)=11.2$~K  and
$a=3.0$~\AA. The respective spectrum  $E(p)$
is shown in Fig.~5b and is characterized by the following values:
$E_\mathrm{max}=14.28$~K
at $p_\mathrm{max}=1.13 \mbox{ \AA}^{-1}$,
$E_\mathrm{min}=9.8$~K at $p_\mathrm{min}=1.95 \mbox{ \AA}^{-1}$;
$c=2.34 \times 10^4 \mbox{ cm/s}$, the total concentration being
$n=2.18\times10^{22}\mbox{ cm}^{-3}$ and the BEC concentration
$n_0=0.09n$ (which agrees with the experimental data \cite{9}).
Such an agreement should be deemed quite satisfactory,
taking into account the simplified model of pair interaction employed.

It is seen from Fig.~4--5 that the nonmonotonous nature of the
spectrum $E(\mathbf{p})$, including, in particular, the presence of
the ``roton'' minimum, is determined mainly by the momentum
dependences of the functions $\tilde \Psi_{11}(\mathbf{p})$ and
$A_0(p)$, which have deep minima due to the oscillations of the
sign-changing potential $\tilde V(p)$ in the domain
$p<2\pi/a$ (see Fig.~1). The theoretical spectra obtained are
in good agreement with the experimental spectrum of $^4$He
both in the positions and the absolute values of the maximum
and minimum of $E(\mathbf{p})$.

One should stress that at comparatively small values of the
parameter $\alpha$ it is impossible to render the quantity
$\tilde \Psi_{12}(0)$ positive, because repulsion between bosons
prevails over effective attraction when integrating over $p$
in Eq.~(\ref{37p}).
At the same time, because of the BEC density being small
($n_0\ll n)$, the quantity 
$\Sigma_{12}(0)=n_0 \Lambda(0)\tilde V(0) -\vert\Psi_{12}(0)\vert$
becomes negative, which corresponds, according to
(\ref{012}) and (\ref{28p}), to a phonon instability of the
spectrum ($\tilde c^2<0$).

On the other hand, at large enough values of $\alpha$,
the effective attraction $\tilde V(p)<0$ in the region $\pi<p/a<2 \pi$
(Fig.~1) turns out to be so strong that around the negative minimum
of $\tilde V(p)$, the radicand in Eqs.~(\ref{22}) and (\ref{38p})
becomes negative, i.e., the quasiparticle spectrum $E(\mathbf{p})$
becomes absolutely unstable (imaginary) in the region
in question, analogously to the Bogolyubov spectrum with a
nonrenormalized potential in the ``hard spheres'' model
(see Fig.~2). Therefore, the domain of parameters $V_0$ and $\alpha$
in which there exists a quasiparticle spectrum (\ref{38p})
which would be stable at all $\mathbf{p}$ and be in agreement
with experiment, is rather narrow.
It is possible, in principle, to consider the inverse problem and
find the effective ``pseudopotential'' of pair interaction of
particles in the $^4$He Bose liquid, starting from the shape of
the empirical quasiparticle spectrum $E_\mathrm{exp}(p)$,
however this is quite tedious.

\section{The criterion of superfluidity and limiting critical velocities}

We conclude with a brief discussion of the applicability of the Landau
criterion of superfluidity to He~II and of the value of limiting
critical velocity in the absence of quantum vortices in the Bose
liquid where the BEC and PCC coexist.

As was noted in Introduction, the elementary excitation spectrum
$E_\mathrm{exp}(\mathbf{p})$ as observed in neutron scattering experiments
\cite{24}--\cite{27}, leads to a rather overestimated value of the
critical velocity, determined by the roton minimum (in accordance
with the Landau criterion), as compared with the experimentally
measured velocities of destruction of the SF flow.
This has to do with the emergence of quantum vortices and vortex
rings in He~II \cite{23}, however under conditions when creation
and/or motion of those vortices is hindered, the critical velocities
increase sharply \cite{28}, \cite{29}, and at low temperatures
$T_c<1K$ they can assume values comparable with 
$v_c=\min \left[\epsilon(p)/p\right]$ \cite{30}.

One should stress that this situation is quite analogous to the one
with the type I superconductors, in which the critical current
$j_c$ is determined by the condition of creation and pinning
of Abrikosov quantum vortices on the surface of the superconductor
or near various defects of the crystal lattice, whereas the
true maximal value of $j_c$---the so-called decoupling critical
current, which is determined by the process of decay of Cooper
pairs \cite{12},---is much larger and can only be observed in
thin wires, whose thickness is much less than the London depth
of penetration of magnetic field into the superconductor, which
hinders vortex creation.

In the $^4$He Bose liquid at finite temperatures, $T\ne 0$,
together with the spectrum $E_\mathrm{exp}(\mathbf{p})$, which at
$\mathbf{p}\to 0$ corresponds to the first (hydrodynamic) sound with
phase velocity $c_1$, there arises---due to the emergence of
the normal component $\rho_n$ in He~II---the second sound,
whose velocity $c_2\ll c_1$ at $T>1$~K.
It is known \cite{36}, \cite{37} that,
due to heat expansion of liquid $^4$He being weak,
the second sound branch reduces
basically to oscillations of temperature (entropy) without any
noticeable transfer of total mass of the normal ($\rho_n$) and SF
($\rho_s$) components, whose oscillations are in counterphase.
Therefore, the excitations of second sound with energy
$\epsilon_2(p)=c_2p$ cannot be observed in standard neutron
scattering experiments, unlike the first sound
$\epsilon_1(p)=c_1p$, which corresponds to cophase oscillations
of the densities $\rho_s$ and $\rho_n$.
At the same time, in the two-component Bose liquid there may coexist
two different types of acoustic Goldstone excitations, associated,
on the one hand, with spontaneous breakdown of gauge symmetry
due to the degeneracy over the phase of the coherent SF condensate
$\rho_s$ at $T\to 0$, and the breakdown of continuous translational
symmetry, i.e., of the homogeneity of the total density 
$\rho=\rho_n+\rho_s$ (first sound); on the other hand, with
spatially inhomogeneous deviations of temperature $T$ from the
uniform distribution due to the oscillations of the density of the
gas of normal excitations (second sound).
Nevertheless, the second sound excitations do carry over some energy
and therefore have to be taken into account in the determination
of the minimum critical velocity, according to the original concept
of the Landau supefluidity criterion \cite{LLD}, which includes all
the types of excitations in the quantum liquid.

In this respect, one can assume that at those
temperatures where $c_2(T)<v_c\simeq 60 \mbox{ m/s}$, the
maximal allowed critical velocity of the macroscopic SF flow
in He~II in the absence (or with strong pinning) of quantum vortices
cannot exceed a number of the order of the second sound velocity
$c_2(T)$, which tends to zero as $T\to T_\lambda$, together with
the SF component density $\rho_{s}(T)$. Precisely such a situation
is characteristic of superconductors, in which the decoupling
critical current turns into zero in the critical point $T=T_c$
together with the energy gap $\Delta$ in the quasiparticle spectrum. 

Finally, it is worth noting that the coexistence of a weak BEC
and an intensive PCC conserves the integer value of the quantum
of circulation of the SF velocity in the vortices $\kappa =\hbar/m$,
due to the total mutual coherence of those condensates in the
SF component $\rho_s$.
Indeed, a sufficiently strong effective attraction for the screened
Fourier component of the singular ``hard spheres'' potential
provides for the formation of a condensate of bound bosonic pairs
with a positive sign of the pair order parameter $\tilde \Psi_{12}(0)$,
whose phase in that case coincides with the one of the BEC.

\section{Conclusions}

Thus, employing the renormalized field theory for the description
of the SF state of the Bose liquid at $T\to 0$ with account for
a small density of the single-particle BEC allows one to formulate
a self-consistent model of superfluidity, in which the SF component
at $T\to 0$ is a coherent superposition of the single-particle BEC,
suppressed due to interaction, and an intensive PCC, which arises
due to an effective attraction between the bosons in the momentum
space. Such an approach lets one obtain, in the framework of the
``hard spheres'' model, an explicit form of the quasiparticle spectrum,
which, with a suitable choice of parameters, coincides with good
precision with the experimental spectrum of elementary excitations
in $^4$He. 

We are sincerely grateful to P.I.~Fomin for numerous useful discussions.

\section{Appendix A}
In order to calculate the Fourier components of the empirical
potentials presented in the Table, it is necessary, in order to avoid
the divergence at $r\to 0$, to make a cutoff at some distance $r_c$,
representing the diameter of a ``hard core''. For example, for the
Lennard--Jones potential, up to the main terms, one has
$$
U(p)=4 \pi \epsilon \int\limits_{r_c}^{\infty}
r dr \sin (pr) \left[\left( \frac{\sigma}{ r }\right)^{12}
-\left( \frac{\sigma}{ r }\right)^{6}\right]\simeq $$
$$\simeq 2 \pi \epsilon \sigma^3
\left[
\frac{1}{5}\left( \frac{\sigma}{ r_c }\right)^{9}
\left( \frac{\sin(p r_c)}{p r_c}+\frac{\cos(p r_c)}{9}-
\frac{p r_c \sin(p r_c)}{7 r}\right)-\right. $$
$$\left.-\frac{1}{2}\left( \frac{\sigma}{ r_c }\right)^{3}
\left\{ \frac{\sin(p r_c)}{p r_c}+\frac{\cos(p r_c)}{3}-
\frac{p r_c \sin(p r_c)}{6}- \right.\right.  \eqno(A.1)$$
$$\left.\left. -\frac{p^2 r_c^2 \cos(p r_c)}{6}-
\frac{p^3 r_c^3}{6}\left( \frac{\pi}{2}-
\mbox{Si}(p r_c)\right)\right\}\right]\;,
$$
where $\mbox{Si}(x)$ is the integral sine.
This potential can be extended to the region $r<r_c$ of the regularized
Fourier component of the infinite repulsion potential in the ``hard spheres''
model \cite{33}, \cite{34} or of the finite potential of ``semi-transparent''
spheres \cite{Pash}.
Such an approach is analogous to the method of pseudopotential in solid state
theory \cite{psevdo}. 
Sign-changing oscillations of the Fourier component (A.1) are associated
with a sharp fracture and a finite jump of the potential at the point $r=r_c$.
In particular, Eq.~(A.1) by itself corresponds, in coordinate space, to
the model of pseudopotential with an ``empty skeleton'', when the potential
is $U(r)=0$ at $r<r_c$, and adding the Fourier components of the potentials
of ``hard'' or ``semitransparent'' spheres to that equation corresponds to
the model of ``filled skeleton'' \cite{psevdo}.

Oscillating sign-changing Fourier components are also characteristic of
certain types of smooth potentials with inflection points.
As an example, apart from the Fermi function (\ref{F}), there is
a potential in the form of a static Lindhardt function
\cite{12} in real space:
$$
W(r)=\frac{W_0}{2}\left[1+\frac{1-x^2}{2x} \log \left\vert \frac{1+x}{1-x}
\right\vert\right]; \qquad x=r/r_0\;, \eqno(A.2)
$$
which has an inflection point with an infinite negative derivative at 
$r=r_0$. The Fourier component of this function can be calculated exactly,
and equals
$$
W(p)=W_0\frac{j_1(p r_0)}{p r_0}\;, \eqno(A.3)
$$
i.e., its momentum dependence coincides with the one of the Fourier
component of the potential of semi-transparent spheres with a finite
jump at $r=r_0$ (see (\ref{F1})).
This is formally equivalent to the calculation of the 
Ruderman-Kittel-Ksui-Yoshida static oscillations \cite{Wite}
in real space for indirect exchange interaction of spin in metals with
the Fermi distribution function at $T=0$ and with the Fourier component
of spin susceptibility of electrons in the form of Eq.~(A.2) with
$x=p/(2 k_F)$ ($k_F$ being the Fourier momentum of the electrons).

At the same time, Fourier components of smooth continuous potentials
with finite derivatives, nonsingular at $r=0$, are characterized by
weak oscillations. Consider for example the modified nonsingular
Buckingham potential (see Table), in which the divergent
term $(-1.49/r^6)$ is replaced, at $r<r_0=2.61\mbox{ \AA}$, with
an exponential $(-B e^{-\beta r})$, finite at $r=0$. The parameters,
$B=0.57 \mbox{ eV}$ and $\beta=1.84 \mbox{ \AA}^{-1}$, are chosen
in such a way that at $r<r_0$, the expression 
$$
V_1(r)=A e^{-\alpha r} -B e^{-\beta r} \eqno(A.4)$$
with $A=481 \mbox{ eV}$,  $\alpha=4.6 \mbox{ \AA}^{-1}$,
would vanish at $r=r_0$, and its derivative over $r$ at that
point would be equal to that of the potential in the $r>r_0$ range:
$$
V_2(r)=A e^{-\gamma r}  -\frac{D}{r^6}-\frac{F}{r^8}\;,
$$
at $C=611 \mbox{ eV}$,
$D=0.94\mbox{ eV \AA}^6$, $F=1.87\mbox{ eV AA}^8$,
and $\gamma=\alpha$.
The Fourier component of this potential is
$$
V(p)=4 \pi \left\{ \frac{2\alpha A}{\left(\alpha^2+p^2 \right)^2}-
\frac{ (A-C)e^{-\alpha r_0}}{p\left(\alpha^2+p^2 \right)}
\left[ \alpha r_0 \sin(pr_0) +p r_0 \cos(pr_0) \right.\right. $$
$$ \left.\left. {}-
\frac{1}{\left(\alpha^2+p^2 \right)^2}\left(
(\alpha^2-p^2) \sin(pr_0)+2\alpha p \cos(pr_0)\right)\right]
-\frac{2\beta B}{\left( \beta^2+p^2\right)^2}\right.$$
$$\left. {}+\frac{B e^{-\beta r_0}}{p\left( \beta^2+p^2\right)}
\left[\beta r_0 \sin(pr_0) +p r_0 \cos(p r_0) \right.\right. \eqno(A.6)$$
$$ \left.\left. {}-\frac{1}{\beta^2+p^2}\left((\beta^2-p^2) \sin(p r_0)+
2 p\beta \cos(p r_0) \right)\right]\right\}-
\frac{4 \pi D}{p} I_5(p)-\frac{4 \pi F}{p} I_7(p)\;,$$
where
$$
I_5(p)=\int\limits_{r_0}^{\infty} \frac{\sin(pr)}{r^5}dr=
\frac{\sin(pr_0)}{4 r_0^4}\left( 1-\frac{p^2r_0^2}{6}\right)$$
$${}+\frac{p \cos(pr_0)}{12 r_0^3}\left( 1-\frac{p^2 r_0^2}{3}\right)+
\frac{p^4}{36}\left[\frac{\pi}{2}-\mbox{Si}(pr_0)\right]\;; \eqno(A.7)$$

$$
I_7(p)=\int\limits_{r_0}^{\infty} \frac{\sin(pr)}{r^7}dr=
\frac{\sin(pr_0)}{6 r_0^6}\left( 1-\frac{p^2r_0^2}{20}-
\frac{p^4r_0^4}{120}\right)$$
$${}+\frac{p \cos(pr_0)}{30 r_0^5}\left( 1-\frac{p^2 r_0^2}{12}
-\frac{p^4 r_0^4}{24}\right)-
\frac{p^6}{720}\left[\frac{\pi}{2}-\mbox{Si}(pr_0)\right]\;. \eqno(A.8)$$

Evidently, the oscillation amplitudes in this case are exponentially
small as compared to the smooth part of the potential.

\section{Appendix B}
The polarization operator (\ref{19}) can be calculated without account
for the vertex part $\Gamma$, making use of the expressions
(\ref{3})--(\ref{6}) in the form
$$
\Pi(\mathbf{p},\omega)=\int \frac{d^3\mathbf{k}}{(2\pi)^3}\left[I_{11}
(\mathbf{p}, \mathbf{k}, \omega)+I_{12}(\mathbf{p}, \mathbf{k}, \omega)
\right]\;,
\eqno(B.1)$$
where
$$
I_{ij}(\mathbf{p}, \mathbf{k}, \omega)=
i\oint \frac{dz}{2\pi} \tilde{G}_{ij}(\mathbf{k}, z)
\tilde{G}_{ij}(\mathbf{k}-\mathbf{p}, z-\omega)\;.
\eqno(B.2)$$
Assume that the Green functions $\tilde{G}_{ij}$ have only one pole
within the integration contour and are equal to
$$
\displaystyle
\tilde{G}_{11} (\mathbf{k}, \epsilon) =
\frac{\epsilon+\frac{k^2}{2m}-\mu+\tilde{\Sigma}_{11}(-\mathbf{k}, -\epsilon)}
{\epsilon^2-E^2(\mathbf{k})+i\delta}\;;
\eqno(B.3)$$
$$
\displaystyle
\tilde{G}_{12} (\mathbf{k}, \epsilon) =
\frac{\tilde{\Sigma}_{12}(\mathbf{k}, \epsilon)}
{\epsilon^2-E^2(\mathbf{k})+i\delta}\;;  \qquad (\delta\to 0)\;.
\eqno(B.4)$$

Calculating the integrals (B2) with account for the poles at the points
$\epsilon=E(\mathbf{k})$ and $\epsilon=E(\mathbf{k}-\mathbf{p})+\omega$ yields
$$\begin{array}{c}
\displaystyle
I_{11}(\mathbf{p}, \mathbf{k}, \omega)=
\frac{1}{2\left[
E(\mathbf{k})-E(\mathbf{k}-\mathbf{p})-\omega\right]}
\left\{
\left[E(\mathbf{k})+\frac{\mathbf{k}^2}{2m}-\mu+\tilde{\Sigma}_{11}
(\mathbf{k}, E(\mathbf{k}))\right]\right.
\\[20pt]
\displaystyle
\left.{}\times\frac{
\left[
E(\mathbf{k})-\omega+\frac{(\mathbf{k}-\mathbf{p})^2}{2m}-\mu+\tilde{\Sigma}_{11}
(\mathbf{k}-\mathbf{p}, E(\mathbf{k})-\omega)\right]}
{E(\mathbf{k})\left[
E(\mathbf{k})+E(\mathbf{k}-\mathbf{p})-\omega\right]}
\right.\\[20pt]
\displaystyle
\left.{}-\left[
E(\mathbf{k}-\mathbf{p})+\frac{(\mathbf{k}-\mathbf{p})^2}{2m}-\mu+\tilde{\Sigma}_{11}
(\mathbf{k}-\mathbf{p}, E(\mathbf{k}-\mathbf{p}))
\right]\right.
\\[20pt]
\displaystyle
\left.{}\times
 \frac{\left[
E(\mathbf{k}-\mathbf{p})+\omega+\frac{\mathbf{k}^2}{2m}
-\mu+\tilde{\Sigma}_{11}
(\mathbf{k}, E(\mathbf{k}-\mathbf{p})+\omega)\right]}
{E(\mathbf{k}-\mathbf{p})\left[
E(\mathbf{k})+E(\mathbf{k}-\mathbf{p})+\omega\right]}\right\}\;,
\end{array}
\eqno(B.5) $$

$$
\begin{array}{c}
\displaystyle
I_{12}(\mathbf{p}, \mathbf{k}, \omega)=
\frac{1}{2\left[
E(\mathbf{k})-E(\mathbf{k}-\mathbf{p})-\omega\right]}
\left\{
\frac{\tilde \Sigma_{12}(\mathbf{k},E(\mathbf{k}))
\tilde \Sigma_{12}(\mathbf{k}-\mathbf{p},E(\mathbf{k})-\omega)}
{E(\mathbf{k})\left[
E(\mathbf{k})+E(\mathbf{k}-\mathbf{p})-\omega\right]}
\right.\\[20pt]
\displaystyle
\left.{}-
\frac{\tilde \Sigma_{12}(\mathbf{k}, E(\mathbf{k}-\mathbf{p})+\omega)
\tilde \Sigma_{12}(\mathbf{k}-\mathbf{p}, E(\mathbf{k}-\mathbf{p}))}
{E(\mathbf{k}-\mathbf{p})\left[E(\mathbf{k})+E(\mathbf{k}-\mathbf{p})+\omega\right]}
\right\}
\end{array}
\eqno(B.6)$$

In the statistical limit $(\omega\to 0,\mathbf{p}\to 0)$, expression (B5)
reduces to
$$
\begin{array}{c}
\displaystyle
I_{11}(0,\mathbf{k},0)=-\frac{1}{4}\left\{\frac{1}{E^2(\mathbf{k})}\left[
\epsilon(k) +\frac{\mathbf{k}^2}{2m}-\mu+
\tilde \Sigma_{11}(\mathbf{k}, \epsilon(k))\right]^2\right.
\\[12pt]
\displaystyle
\left.
{}+\left[
\frac{2}{\epsilon(k)}\left(1+\frac{\partial \tilde \Sigma_{11}(\mathbf{k})}
{\partial \epsilon}\right)
-\frac{k}{m\epsilon(k)}
\frac{1}{\frac{\partial\epsilon(k) }{\partial k}}\right]
\left[
\epsilon(k) +\frac{\mathbf{k}^2}{2m}-\mu+
\tilde \Sigma_{11}(\mathbf{k}, \epsilon(k))\right]\right\}
\end{array}
\eqno(B.7)$$
It follows that in a large region of momentum space,
$I_{11}(0,\mathbf{k},0)<0$. The same result is obtained for the
function (B.6) at $p=0$ and $\omega=0$, i.e., $I_{12}(0,\mathbf{k},0)<0$,
so that the static bosonic polarization operator $\Pi(0,0)$ is negative,
which corresponds to a suppression of the ``screened'' repulsion at
$\mathbf{p}\to 0$. From Eqs.~(B.5) and (B.6) it can also be seen that on
the mass shell $\omega=E(\mathbf{p})$, the integrals $I_{11}$ and $I_{12}$
remain negative in a wide region of momentum space because of the
negative sign of the common denominator
$\left[E(\mathbf{k})-E(\mathbf{k}-\mathbf{p})-E(\mathbf{p})\right]<0$
and the positive sign of the denominator
$\left[E(\mathbf{k})+E(\mathbf{k}-\mathbf{p})-E(\mathbf{p})\right]>0$
due to the fact that the quasiparticle spectrum $E(\mathbf{p})$ is decayless
[conditions (\ref{34p})].

Thus, the polarization operator on the mass shell $\epsilon=E(\mathbf{p})$
is negative on the whole range $p<2\pi/a$, where $\sin(pa)<0$.
It is necessary to stress that this feature of the polarization operator,
$\Pi (\mathbf{p}, E(p))<0$, is only characteristic of Bose systems, in which
the single-particle and collective spectra coincide with each other
and are counted from the
common zero of energy, unlike the Fermi systems, in which the single-particle
excitation spectrum begins at the Fermi energy, due to the Pauli principle.
Therefore, in the Fermi liquid ($^3$He)
there can be no corresponding effective
enhancement of the negative values
of the ``input'' interaction potential $V(p)$,
so that the formation of Cooper pairs is only possible for nonzero orbital
momenta, due to the true weak van der Waals attraction between fermions.
Apparently, it is this fact that has to do with the critical temperatures
of the SF transition in $^4$He i $^3$He
differing by three orders of magnitude.

\newpage
\begin{tabular}{|p{2.5cm}|p{9.5cm}|p{3.0cm}|}
\hline
Name of potential & Form of potential $\Phi(r)$ & Energy level\\
\hline
\hline
  Rosen-Margenau \cite{T3},\cite{T4}, \cite{T5}&
$
\lbrack 925 \exp(-4.4 r) -560\exp (-5.33 r)-\frac{1.39}{r^6}
-\frac{3}{r^8}\rbrack \cdot 10^{-12} \;\mbox{ erg} $ & None\\
\hline
 Slater-Kirkwood \cite{T5}, \cite{T6}&
$\lbrack 770 \exp(-4.6 r) -\frac{1.49}{r^6}
\rbrack \cdot 10^{-12} \;\mbox{ erg}$ & None\\
\hline
 Intem-Schneider \cite{T2}& $\lbrack 1200 \exp(-4.72 r) -\frac{1.24}{r^6}
-\frac{1.89}{r^8}\rbrack \cdot 10^{-12} \;\mbox{ erg}$ & None\\
\hline
 &
$ 4\epsilon\left[
\left(\frac{\sigma}{r}\right)^{12}-
\left(\frac{\sigma}{r}\right)^{6}\right] $ & \\
Lennard&
Case 1 \cite{T1}:
$ \sigma=2.556, \epsilon=10.22\;\mbox{ K}$& None\\
-Jones &
Case 2 \cite{enciklop}:
$ \sigma=2.642, \epsilon=10.80\;\mbox {K}$.&
$\epsilon=0.0108\;\mbox{ K}$\\
\hline
Buckingham \cite{T7}&
$\left\{\begin{array}{ll}
\lbrack 770 \exp(-4.6 r)-\frac{1.49}{r^6}
\rbrack  10^{-12} , \quad  r\leq 2.61 \;\mbox{ \AA};\\ \\
\lbrack 977 \exp(-4.6 r)-\frac{1.50}{r^6} -\frac{2.51}{r^8}
\rbrack  10^{-12},  r\geq 2.61 \;\mbox{ \AA}.
\end{array}\right. $&
$\varepsilon=0.00632\;\mbox{ K}$\\
\hline
Massey-Buckingham \cite{T7} &
$\lbrack 1000 \exp(-4.6 r) -\frac{1.91}{r^6}
\rbrack \cdot 10^{-12} \;\mbox{ erg}\;$&
$\varepsilon=0.0622\;\mbox {K}$\\
\hline
 Buckingham-Hamilton \cite{T7}&
$\lbrack 977 \exp(-4.6 r) - \frac{1.5}{r^6}
-\frac{2.51}{r^8}\rbrack \cdot 10^{-12} \;\mbox{ erg}\; $ &
$\varepsilon=0.00229\;\mbox{ K}$\\
\hline
\end{tabular}

\newpage
{\large\bf Figure Captions}

Fig.~1. { The dependence of the ratio
$\displaystyle\frac{\tilde V(pa)}{V_0 }$
on the dimensionless momentum $p$ for different values of the
dimensionless parameter
$\alpha=V_0\vert\overline {\Pi}\vert$: Dashed curve 1,
the ``hard spheres'' input potential at $\alpha$=0;
solid curve 2 corresponds to $\alpha$=2,
curve 3---to $\alpha$=3; curve 4---to $\alpha$=3.5.}

Fig.~2. {The Bogolyubov spectrum (\ref{32p})
for a dilute quasi-ideal Bose gas, obtained by substituting
the potential (\ref{31p}) with an independent choice
of the two parameters $\alpha=2.5$ and
$V_0/(4\pi a^3)=169.9\mbox{ K}$ at $a=2.44\mbox{ \AA}$.
The solid curve corresponds to the instability of the
Bogolyubov spectrum obtained by substituting the potential
(\ref{31p}) into (\ref{32p}) for
$n=2.17 \times10^{22} \mbox{ cm}^{-3}$ i $a=2.44\mbox{ \AA}$,
characteristic of $^4$He.}

Fig.~3. (a) {The numerators $F_+$ i $F_-$
of both terms in expressions (\ref{PI2}) and (\ref{PI3})
remain positive at any $\mathbf{p}$ and $\mathbf{k}$.}
(b) {The $\mathbf{p}$ dependence of the polarization
operator $\Pi(\mathbf{p}, E(\mathbf{p}))$ (\ref{PI1})
on the mass shell, obtained numerically by iterations.}

Fig.~4. {The momentum dependence of the functions (a) $\Psi_{11}(p)$,
(b) $\Psi_{12}(p)$, (c) $A_0(p)$, obtained according to
Eqs.~(\ref{36p}), (\ref{37p}) i (\ref{39p}), for
$V_0/(4\pi^2 a^3)=17.7\mbox{ K}$,\quad$\alpha=4.4$.}

Fig.~5. {(a) The elementary excitation spectrum $E(p)$ obtained
from Eq.~(\ref{39p}), for the same set of parameters as on Fig.~4.
The maximum energy in the quasiparticle spectrum (\ref{38p}) at
$p/\hbar=1.05\mbox{ \AA}^{-1}$ is $E_\mathrm{max}=15.9\mbox{ K}$,
and the minimum one, at the roton minimum at
$p/\hbar=1.72\mbox{ \AA}^{-1}$, is $E_\mathrm{min}=10.7\mbox{ K}$.
The hydrodynamic sound velocity
$c_1=2.08\times10^4 \mbox{ cm/s}$ and the total quasiparticle
concentration $n=2.57\times10^{22} 1/\mbox{cm}^3$
at the BEC concentration $n_0=0.03n$ and the diameter
$a=2.44\mbox{ \AA}$, found according to Eqs.~(\ref{28p}) and
(\ref{25a});
(b) The spectrum $E(p)$ for $V_0/(4\pi^2 a^3)=11.2\mbox{ K}$,\quad
$\alpha=4.52$. The maximum energy of the spectrum (\ref{38p}) at
$p/\hbar=1.23\mbox{ \AA}^{-1}$ is $E_\mathrm{max}=15.85\mbox{ K}$,
and the minimum one, at the roton minimum at
$p/\hbar=2\mbox{ \AA}^{-1}$ is $E_{min}=9.81\mbox{ K}$.
The hydrodynamic sound velocity
$c_1=2.34\times10^4 \mbox{ cm/s}$ and the total quasiparticle
concentration $n=2.18\times10^{22} 1/\mbox{ cm}^3$
at the BEC concentration $n_0=0.09n$ and the diameter
$a=2.95\mbox{ \AA}$, found according to(\ref{28p}) and
(\ref{25a}).
The dashed line corresponds to the experimental curve.}

\newpage
\begin{center}
\includegraphics[57,3][361,191]{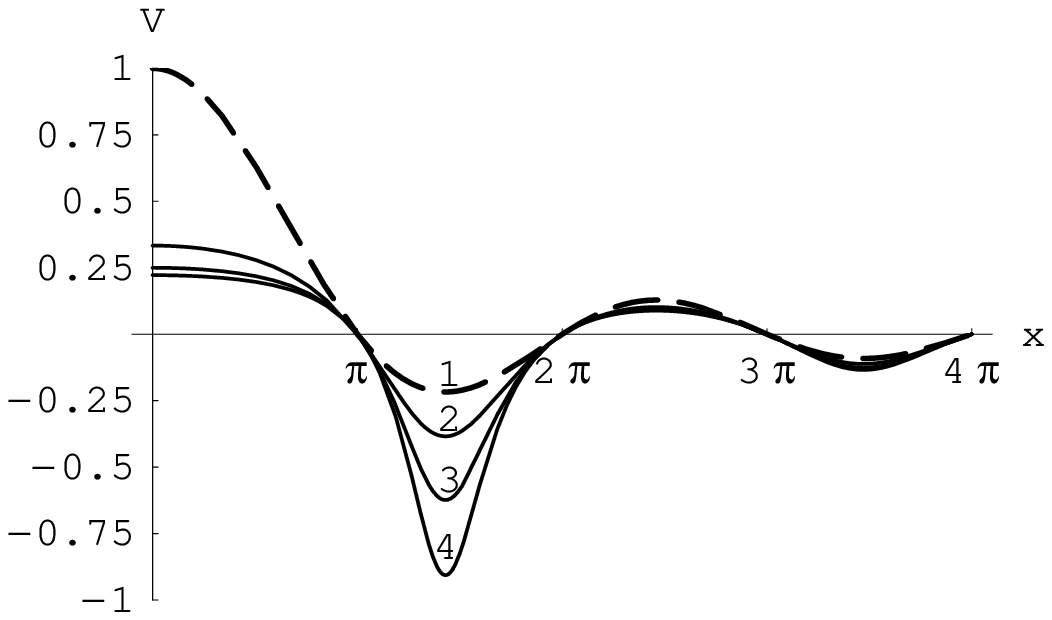} \\
Fig.~1
\end{center}

\newpage
\begin{center}
\includegraphics[14,14][459,356]{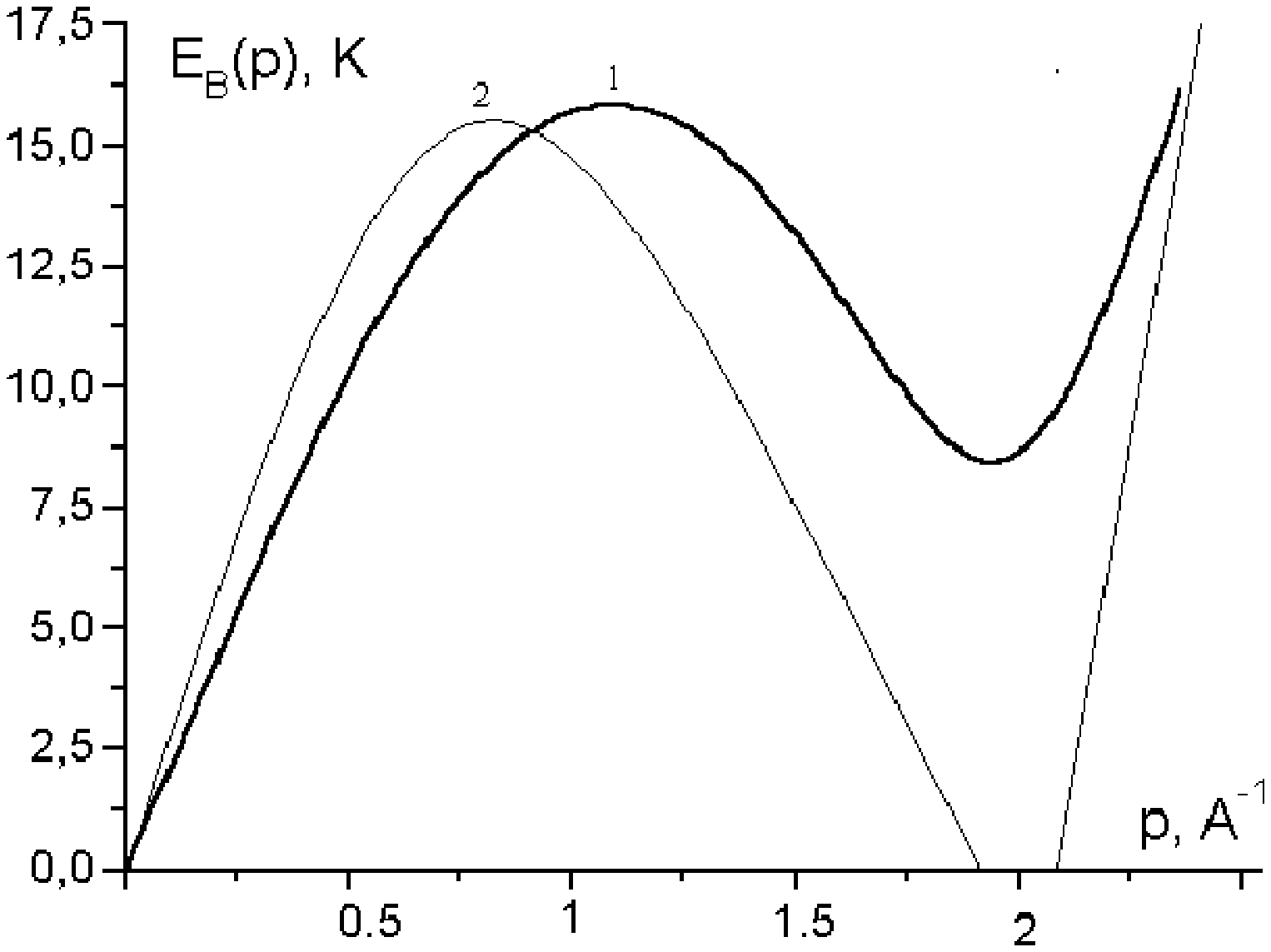} \\
Fig.~2
\end{center}

\newpage
\begin{center}
\includegraphics[14,14][386,253]{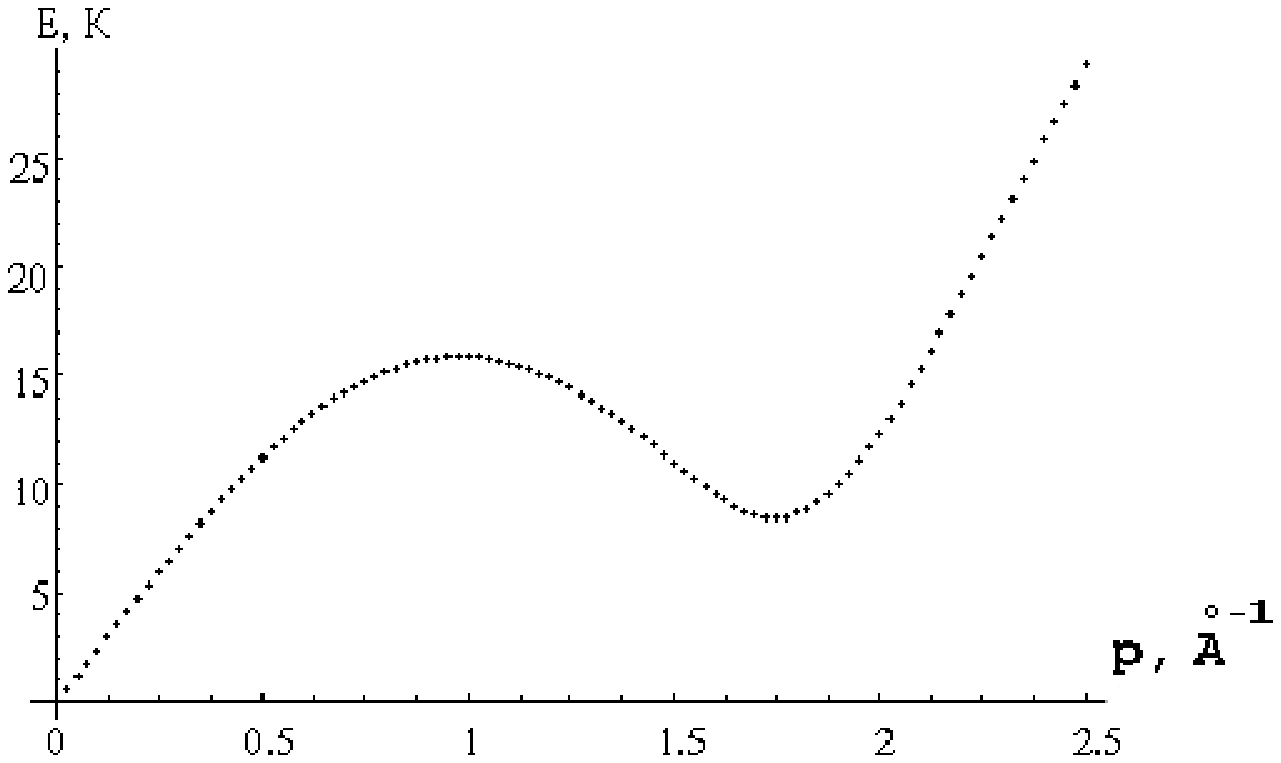} \\
Fig.~3a
\end{center}

\newpage
\begin{center}
\includegraphics[14,14][379,242]{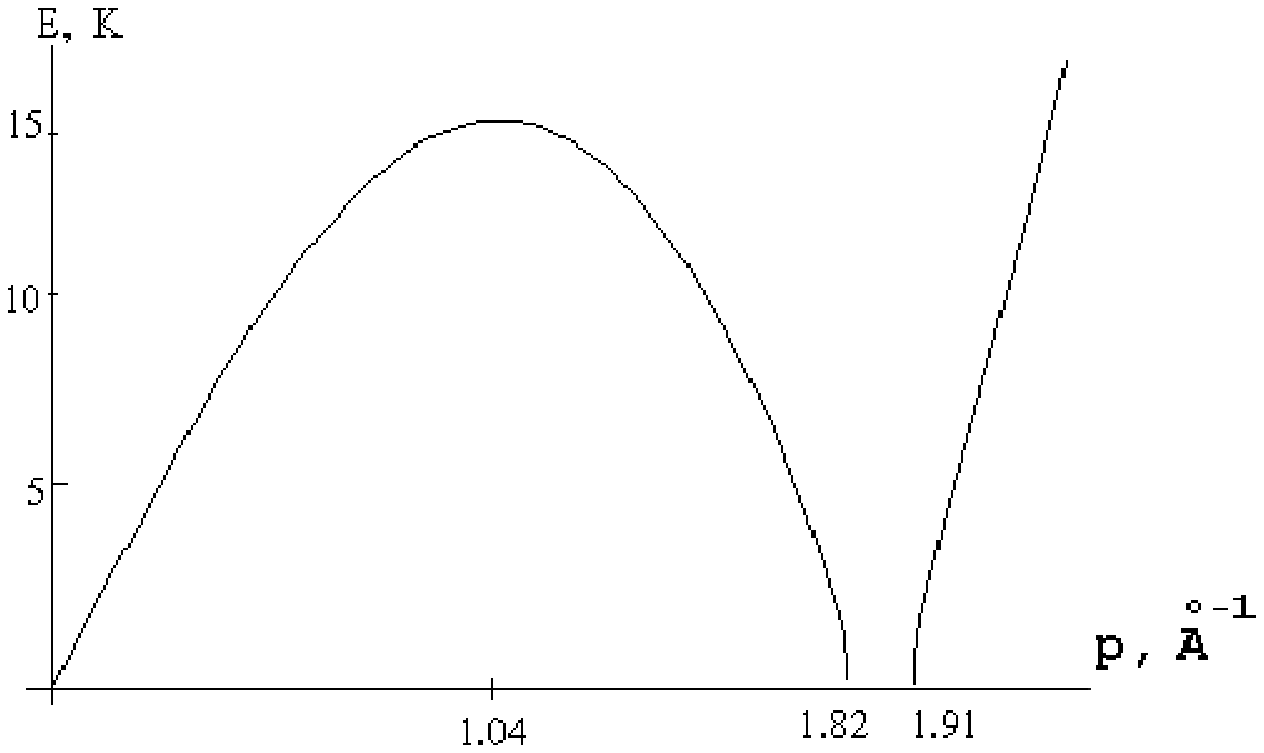} \\
Fig.~3b
\end{center}

\newpage
\begin{center}
\scalebox{0.7}{\includegraphics[14,14][527,365]{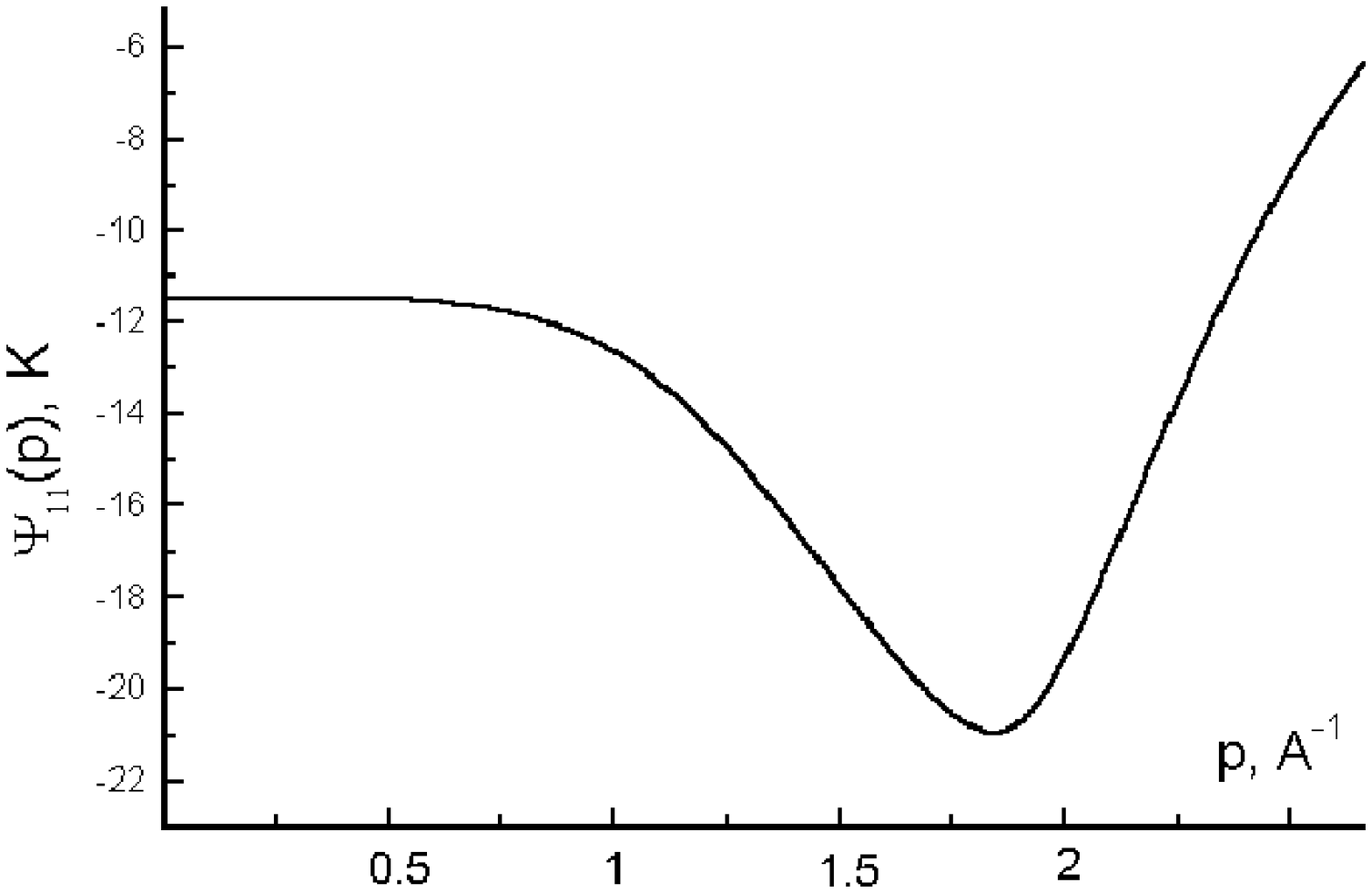}} \\
Fig.~4a
\end{center}

\newpage
\begin{center}
\scalebox{0.7}{\includegraphics[14,14][526,354]{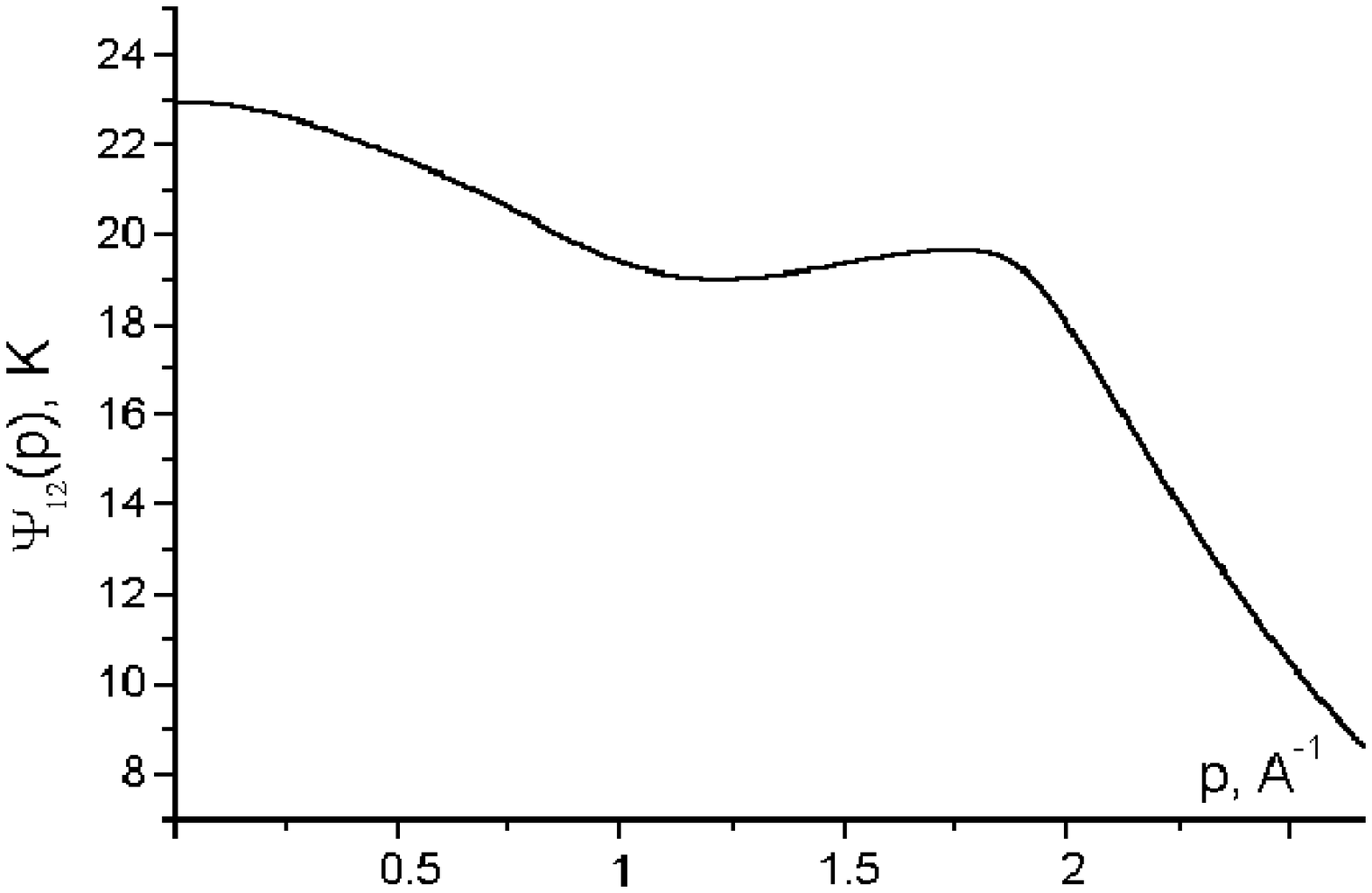}} \\
Fig.~4b
\end{center}

\newpage
\begin{center}
\scalebox{0.7}{\includegraphics[14,14][485,358]{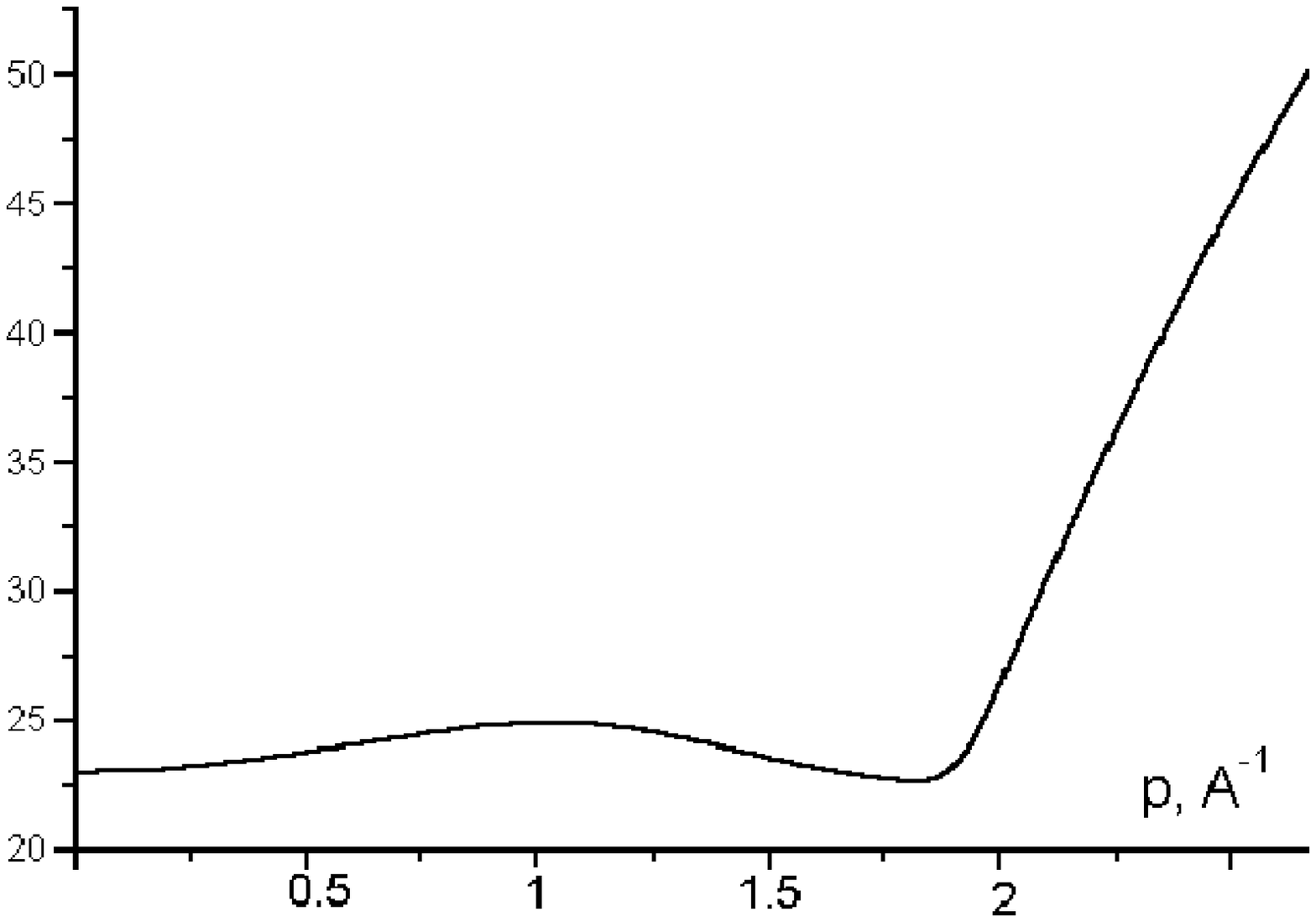}} \\
Fig.~4c
\end{center}

\newpage
\begin{center}
\scalebox{0.7}{\includegraphics[14,14][532,359]{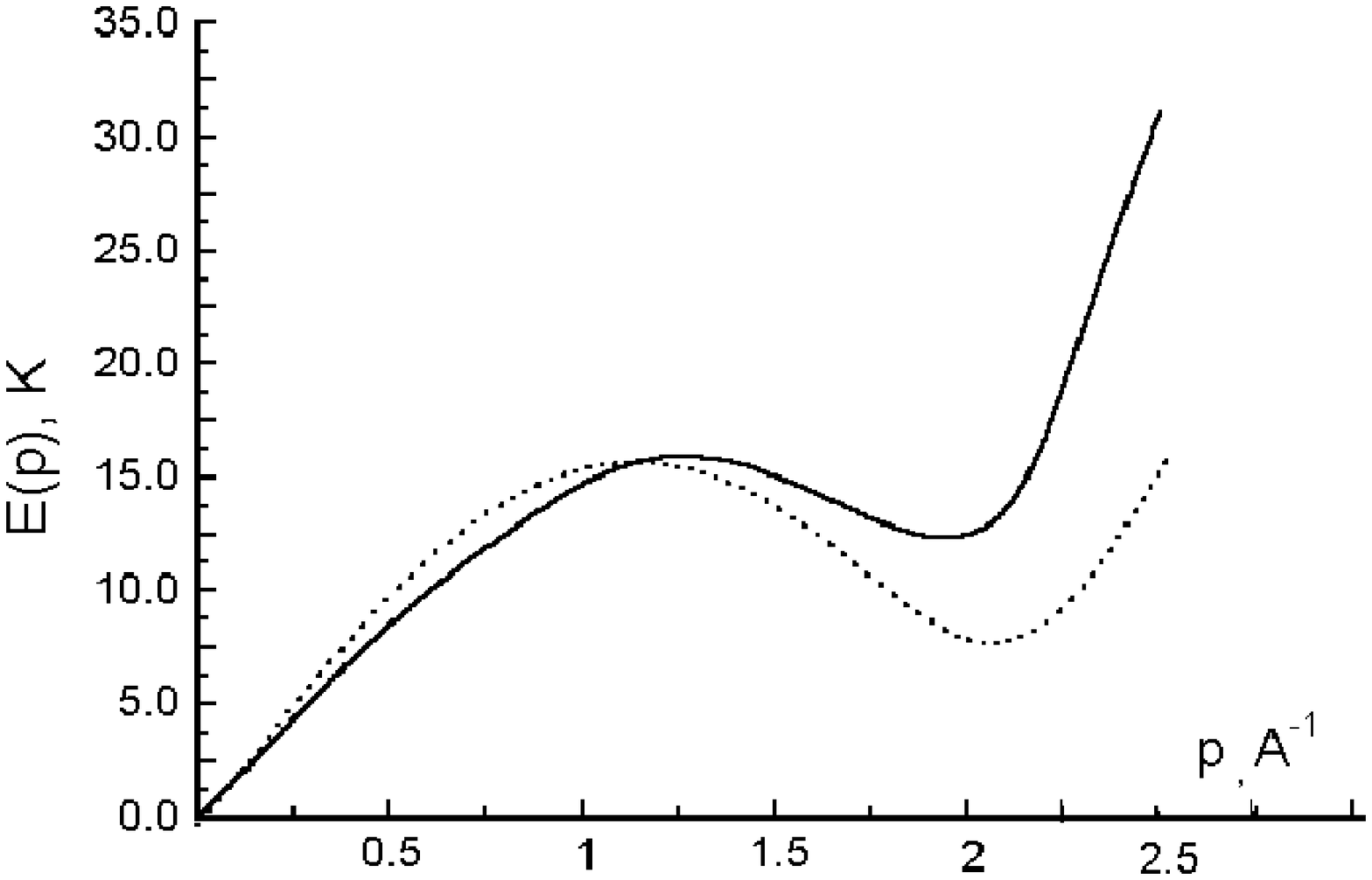}} \\
Fig.~5a
\end{center}

\newpage
\begin{center}
\scalebox{0.7}{\includegraphics[14,14][540,360]{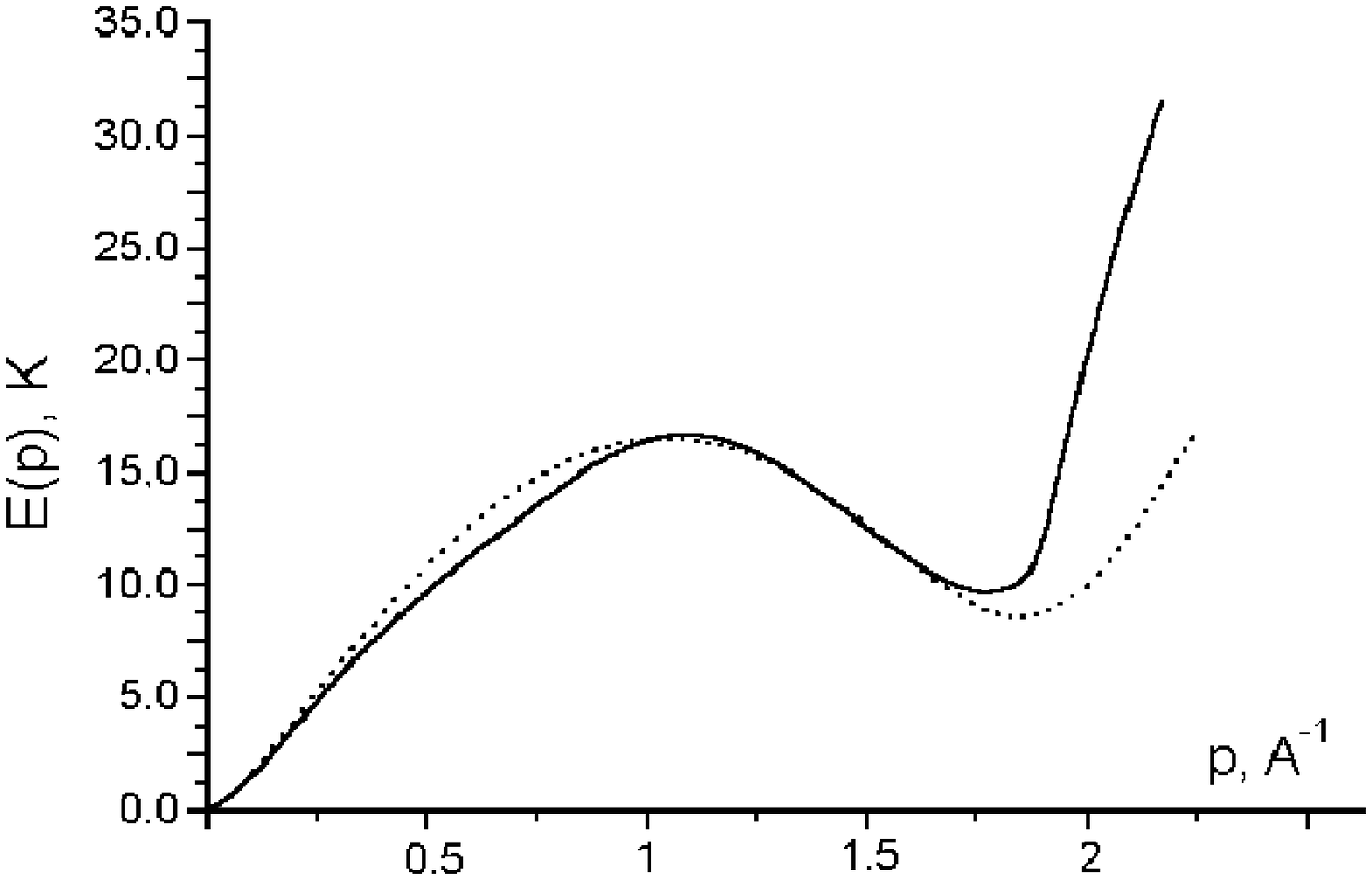}} \\
Fig.~5b
\end{center}

\end{document}